\documentclass[useAMS,usenatbib]{mn2e}

\usepackage{graphicx} 

\title[X-ray outburst in 3C~111]{An outburst scenario for the X-ray spectral variability in 3C~111}
\author[F. Tombesi et al.]{F. Tombesi$^{1,2}$\thanks{E-mail: ftombesi@astro.umd.edu}, J. N. Reeves$^{3}$, C. S. Reynolds$^{2}$, J. Garc{\'{\i}}a$^{2,4}$ and A. Lohfink$^{2}$\\  
$^{1}$X-ray Astrophysics Laboratory, NASA/Goddard Space Flight Center, Greenbelt, MD 20771, USA\\
$^{2}$Department of Astronomy, University of Maryland, College Park, MD 20742, USA\\
$^{3}$Astrophysics Group, School of Physical and Geographical Sciences, Keele University, Keele, Staffordshire ST5 5BG, UK\\
$^{4}$Harvard-Smithsonian Center for Astrophysics, 60 Garden St., Cambridge, MA 02138 USA\\
}
\begin{document}

\date{Accepted 2013 July 1. Received 2013 June 26; in original form 2013 April 19.}


\maketitle

\label{firstpage}

\begin{abstract}

We present a combined \emph{Suzaku} and \emph{Swift} BAT broad-band E$=$0.6--200~keV spectral analysis of three 3C~111 observations obtained in 2010. The data are well described with an absorbed power-law continuum and a weak ($R\simeq 0.2$) cold reflection component from distant material. We constrain the continuum cutoff at $E_\mathrm{C}$$\simeq$150--200~keV, which is in accordance with X-ray Comptonization corona models and supports claims that the jet emission is only dominant at much higher energies. Fe~XXVI Ly$\alpha$ emission and absorption lines are also present in the first and second observations, respectively. The modelling and interpretation of the emission line is complex and we explore three possibilities. If originating from ionized disc reflection, this should be emitted at $r_\mathrm{in} \ge 50$~$r_g$ or, in the lamp-post configuration, the illuminating source should be at a height of $h \ge 30$~$r_g$ over the black hole. Alternatively, the line could be modeled with a hot collisionally ionized plasma with temperature $kT = 22.0^{+6.1}_{-3.2}$~keV or a photo-ionized plasma with log$\xi$$=$$4.52^{+0.10}_{-0.16}$~erg~s$^{-1}$~cm and column density $N_\mathrm{H} > 3\times 10^{23}$~cm$^{-2}$. However, the first and second scenarios are less favored on statistical and physical grounds, respectively. The blue-shifted absorption line in the second observation can be modelled as an ultra-fast outflow (UFO) with ionization parameter log$\xi$$=$$4.47^{+0.76}_{-0.04}$~erg~s$^{-1}$~cm, column density $N_\mathrm{H}$$=$$(5.3^{+1.8}_{-1.3})\times 10^{22}$~cm$^{-2}$ and outflow velocity $v_\mathrm{out} = 0.104\pm0.006$c. 
Interestingly, the parameters of the photo-ionized emission model remarkably match those of the absorbing UFO, supporting the possibility that the same material could be responsible for both emission and absorption. We suggest an outburst scenario in which an accretion disc wind, initially lying out of the line of sight and observed in emission, then crosses our view to the source and it is observed in absorption as a mildly-relativistic UFO.

\end{abstract}

\begin{keywords}
accretion, accretion discs -- black hole physics -- line: identification -- plasmas -- galaxies: active -- X-rays: galaxies
\end{keywords}

\section{Introduction}

In radio-loud Active Galactic Nuclei (AGNs) highly collimated relativistic jets are routinely observed at radio, optical and X-rays. They are thought to be powered by the extraction of rotational energy from the spinning super-massive black hole (SMBH) and they can travel for very long distances, impacting areas far away from the center of these galaxies. Until very recently, this was the main known mechanism for the deposition of mechanical energy from the SMBH into the host galaxy environment in radio-loud sources, contributing to the AGN cosmological feedback (e.g., Fabian 2012). Deep X-ray observations changed this view, showing an increasing evidence for accretion disc outflows in this class of sources as well. 

For instance, a systematic analysis of the \emph{Suzaku} spectra of five bright Broad-Line Radio Galaxies (BLRGs) in the E$=$4--10~keV band showed significant blue-shifted Fe XXV/XXVI K-shell absorption lines at E$>$7~keV in three sources (Tombesi et al.~2010a), namely 3C~111, 3C~120, 3C~390.3. They imply an origin from highly ionized (log$\xi$$\sim$3--6~erg~s$^{-1}$~cm) and massive ($N_H$$\sim$$10^{22}$--$10^{24}$~cm$^{-2}$) gas outflowing with mildly relativistic velocities of $v_{out}$$\sim$0.1c.

These characteristics are very similar to those of the Ultra-Fast Outflows (UFOs) with velocities $v_{out} \ge$10,000~km/s observed in $\ga$40\% of radio-quiet AGNs, in particular local Seyferts and quasars (e.g., Chartas et al. 2002, 2003; Pounds et al.~2003; Dadina et al.~2005; Markowitz et al.~2006; Papadakis et al.~2007; Braito et al.~2007; Cappi et al.~2009; Reeves et al.~2009; Chartas et al.~2009; Giustini et al.~2011; Gofford et al.~2011; Lobban et al.~2011; Patrick et al.~2012; Dauser et al.~2012; Tombesi et al.~2010b, 2011a; Gofford et al.~2013). Therefore, we might be witnessing the same phenomena in both radio-quiet/loud AGNs.

The mechanical (or kinetic) power of UFOs is rather high, in the range log$\dot{E}_\mathrm{K}$$\simeq$42--45~erg~s$^{-1}$ (Tombesi et al.~2012a; 2013), a value that is systematically higher than $\sim$0.1--0.5\% of the AGN bolometric luminosity, the minimum required by simulations of feedback induced by winds/outflows (e.g. Hopkins \& Elvis 2010; Gaspari et al.~2011a, 2011b, 2012b). Therefore, in the long term, they could be able to significantly influence SMBH growth and the observed black hole-host galaxy relations, such as the $M_{BH}$--$\sigma$ (Ferrarese \& Merritt 2000; King 2010; Ostriker et al.~2010; Gaspari et al.~2011a, 2011b, 2012a, 2012b; Wagner et al.~2013). 

The UFOs are likely launched from the inner accretion disc and the observed parameters and correlations are, to date, consistent with both radiation pressure through Compton scattering and magnetohydrodynamic (MHD) processes contributing to the outflow acceleration (e.g., King \& Pounds 2003; Proga \& Kallman 2004; Everett \& Ballantyne 2004; Everett 2005; Ohsuga et al.~2009; Fukumura et al.~2010; Kazanas et al.~2012). The study of disc winds/outflows in BLRGs represents an important step for improving our understanding of the jet-disc connection and models attempting to explain this phenomenology should take these components into account (e.g., McKinney~2006; Yuan, Bu \& Wu 2012). 

3C~111 is a powerful and X-ray bright radio galaxy at $z$=0.0485 with a SMBH mass of log$M_{\mathrm{BH}}$$=$$8.1\pm0.5$~$M_{\odot}$ (Chatterjee et al.~2011). The inclination of the radio jet is $\theta = 18^{\circ}\pm5^{\circ}$ (Jorstad et al.~2005) or 10$^{\circ} < \theta < 26^{\circ}$ (Lewis et al.~2005). 
The source has been recently the subject of an extensive monitoring campaign to study its accretion disc-jet connection (Chatterjee et al.~2011). Indeed, major X-ray dips in the light curve are followed by ejections of bright superluminal knots in the radio jet. 3C~120 also shows a similar behaviour (Marscher et al.~2002; Chatterjee et al.~2009) and it might have analogies with stellar-mass black holes (e.g., Neilsen \& Lee 2009; Fender et al.~2009).

In this respect, Tombesi et al.~(2012b) presented a comparison of the parameters of the UFOs and the jet of 3C~111 on sub-pc scales. They find that the superluminal jet coexists with mildly relativistic outflows on sub-pc scales and that both of them are powerful enough to exert a concurrent feedback impact on the surrounding environment. There are also evidences suggesting that UFOs might be placed within the known X-ray dips-jet ejection cycles as well. In this paper we exploit the unique capabilities of the combined instruments on board \emph{Suzaku} and \emph{Swift} BAT to perform a broad-band spectral analysis of 3C~111 in the interval E$=$0.6--200~keV. This is complementary to the E$=$4--10~keV study performed in Tombesi et al.~(2011b). Here, the broad-band spectral analysis is mainly required to constrain the overall continuum and self-consistently constrain the parameters of neutral/ionized reflection models, including both Fe K emission lines and the high-energy reflection hump.

\section{Data reduction}

The details of the three September 2010 \emph{Suzaku} observations are summarized in Table~1. They are spaced by $\sim$7 days and we refer to them as Obs1, Obs2 and Obs3, respectively.

\subsection{\emph{Suzaku} XIS}

\emph{Suzaku} observed 3C~111 three times in September 2010 for a net exposure time after screening of $\sim$60~ks for each case. The observations are spaced by $\sim$7~days. We use the cleaned event files obtained from the \emph{Suzaku} pipeline processing with standard screening criteria applied. The main parameters of the XIS observations are reported in Table~1. The source spectra were extracted from circular regions of 3{\arcmin} radius centered on the source, whereas background spectra were extracted from a region of same size offset from the main target and avoiding the calibration sources. The XIS response files were produced. The spectra from the front illuminated (FI) CCDs, XIS0 and XIS3, were combined after verifying that the data are consistent with each other. The data of the back illuminated (BI) XIS1 will not be used in the following spectral analysis due to the much lower sensitivity and the possible cross-calibration uncertainties with the XIS-FI. However the relative parameters are listed in Table~1 for completeness. The 4--10~keV XIS-FI lightcurve of the three observations was reported in Fig.~1 of Tombesi et al.~(2011b). There is a $\sim$30\% flux increase between the first observation and the last two.

\begin{table}
\begin{minipage}{75mm}
\caption{3C~111 observations log.}
\begin{tabular}{@{\hspace{0.2cm}}l@{\hspace{0.2cm}}c@{\hspace{0.2cm}}c@{\hspace{0.2cm}}c@{\hspace{0.2cm}}c@{\hspace{0.2cm}}c@{\hspace{0.2cm}}}
\hline
Satellite & Obs & Date & Instr & Exp & Rate \\
    &     &      &      & (ks) & (cts/s) \\
\hline
\emph{Suzaku} & 1 & 2010/09/02 & XIS-FI & 59 & 1.246$\pm$0.003 \\
              &   &            & XIS-BI & 59 & 1.167$\pm$0.005 \\
              &   &            & PIN    & 68 & 0.139$\pm$0.003 \\
              &   &            & GSO    & 68 & 0.062$\pm$0.020 \\
              & 2 & 2010/09/09 & XIS-FI & 59 & 1.640$\pm$0.004 \\
              &   &            & XIS-BI & 59 & 1.604$\pm$0.005 \\
              &   &            & PIN    & 67 & 0.166$\pm$0.003 \\
              &   &            & GSO    & 67 & 0.050$\pm$0.020 \\
              & 3 & 2010/09/14 & XIS-FI & 52 & 1.587$\pm$0.004 \\
              &   &            & XIS-BI & 52 & 1.608$\pm$0.006 \\
              &   &            & PIN    & 65 & 0.141$\pm$0.003 \\
              &   &            & GSO    & 68 & 0.055$\pm$0.020 \\
\emph{Swift}  &   & 58-month   & BAT    &    &  0.002\\

\hline
\end{tabular}
{\em Note.} The net source count-rates after background subtraction refer to each instrument separately and are in the 2--10~keV band for the XIS, in the 15--70~keV band for the PIN, in the 70--200~keV band for the GSO and in the 15--200~keV band for the BAT, respectively.
\end{minipage}
\end{table}

\subsection{\emph{Suzaku} HXD and \emph{Swift} BAT}

For the HXD-PIN data reduction and analysis we followed the standard procedures. 
We use the rev2 data, which include all four cluster units and the best background available, which account for both the instrumental and cosmic X-ray components (Kokubun et al. 2007). The source and background spectra were extracted within the common good time intervals and the source spectrum was corrected for the detector dead-time. The latest response file was used. The net exposure times after screening are reported in Table~1.

In the E$=$15--70~keV band, the PIN spectrum can be fitted with a single power law with $\Gamma_{PIN}=$ 1.64$\pm$0.13, 1.57$\pm$0.11 and 1.65$\pm$0.11, yielding a flux of $1.02^{+0.05}_{-0.13}\times 10^{-10}$~erg~s$^{-1}$~cm$^{-2}$, $1.27^{+0.04}_{-0.12}\times 10^{-10}$~erg~s$^{-1}$~cm$^{-2}$  and $1.05^{+0.03}_{-0.15}\times 10^{-10}$~erg~s$^{-1}$~cm$^{-2}$ for the first, second and third observations, respectively (with 90\% errors). The ratio of the source PIN countrate with respect to the total background countrate for the three observations is 45\%, 53\% and 40\%, respectively. This is much higher than the typical accuracy of the PIN background model of less than 3--5\%. 3C~111 is also marginally detected ($\sim 2 \sigma$) with the HXD-GSO, with count-rates reported in Table~1 (corresponding to $\sim$0.5\% of the total). However, given the very low signal-to-noise of these data, the GSO spectra are not further considered in this work.

The \emph{Swift} BAT spectrum was derived from the 58-Month hard X-ray survey\footnote{http://heasarc.gsfc.nasa.gov/docs/swift/results/bs58mon/}. The data reduction and extraction procedure of the spectrum is described in Baumgartner et al. (2010). We use the latest calibration response diagonal\_8.rsp and background files. The source is clearly detected in the energy band E$=$15--200~keV with a signal-to-noise (S/N) of 37 and a count rate of ($1.77\pm0.05)\times 10^{−3}$ cts~s$^{-1}$. The E$=$50--200~keV flux of $\sim$$10^{-11}$~erg~s$^{-1}$~cm$^{-2}$ is consistent with the GSO.

\section{Broad-band spectral analysis}

\subsection{Baseline models}

We consider the three observation separately due to the different flux levels. The spectral analysis was carried out using the \emph{heasoft} v.6.12 package and XSPEC v.12.7.1 (Arnaud 1996). All uncertainties quoted are at 1$\sigma$ level for one parameter of interest and the energy of the spectral lines is reported in the source rest-frame, unless otherwise stated. The spectrum of the \emph{Suzaku} XIS-FI was fitted in the E$=$0.6--10.5~keV band, excluding the energy range around the Si K edge (E$=$1.5--2~keV), which is known to be affected by calibration issues (Koyama et al.~2007). The \emph{Suzaku} PIN and \emph{Swift} BAT data were included in the E$=$15--70~keV and E$=$15--200~keV bands, respectively. All spectra were grouped to a minimum of 25 counts per energy bin in order to allow the use of the $\chi^2$ minimization in the model fitting. 

When fitting the combined broad-band spectrum, a cross-normalization factor of 1.16 between the \emph{Suzaku} XIS/PIN was included. Instead, given that the \emph{Swift} BAT data are averaged over about 5 years, we leave free the XIS/BAT cross-normalization in order to account for the spectral variability (e.g., Lobban et al.~2010). This latter value is found to be 0.85, 0.72 and 0.81 for the three observations, respectively. Throughout the spectral analysis, we always include a neutral Galactic absorption component of $N_\mathrm{H} = 3.0 \times 10^{21}$ cm$^{-2}$ (Kalberla et al.~2005). However, the source is known to show intrinsic neutral absorption in excess of the Galactic one of $N_\mathrm{H}$$\simeq$$6\times 10^{21}$ cm$^{-2}$ and a narrow Fe K$\alpha$ emission line at E$\simeq$6.4~keV (e.g., Reynolds et al.~1998; Ballo et al.~2011; Tombesi et al.~2011b). These components are clearly apparent from the ratios of the spectra with respect to the extrapolated 4--10~keV Galactic absorbed power-law continuum shown in Fig.~1. 

Taking advantage of the broad-band spectra, here we do not model the Fe K$\alpha$ emission line with a simple Gaussian as in Tombesi et al.~(2011b), but we employ the cold reflection model \emph{pexmon} (Nandra et al.~2007). This allows us to self-consistently take into account both the emission line and the reflection hump. We assume standard Solar abundances and an inclination consistent with that of the jet of $\simeq$18$^{\circ}$ (Jorstad et al.~2005). Therefore, the baseline model for these observations consists of a Galactic absorbed power-law continuum with intrinsic neutral absorption and a cold reflection component. As we can see from Fig.~2, this model provides a very good characterization of the spectra, with reduced $\chi^2/d.o.f.$ values of 1.022, 1.033 and 1.028, respectively. The best-fit parameters and statistics are reported in Table~2.

  \begin{figure*}
  \centering
   \includegraphics[width=4.3cm,height=5.5cm,angle=-90]{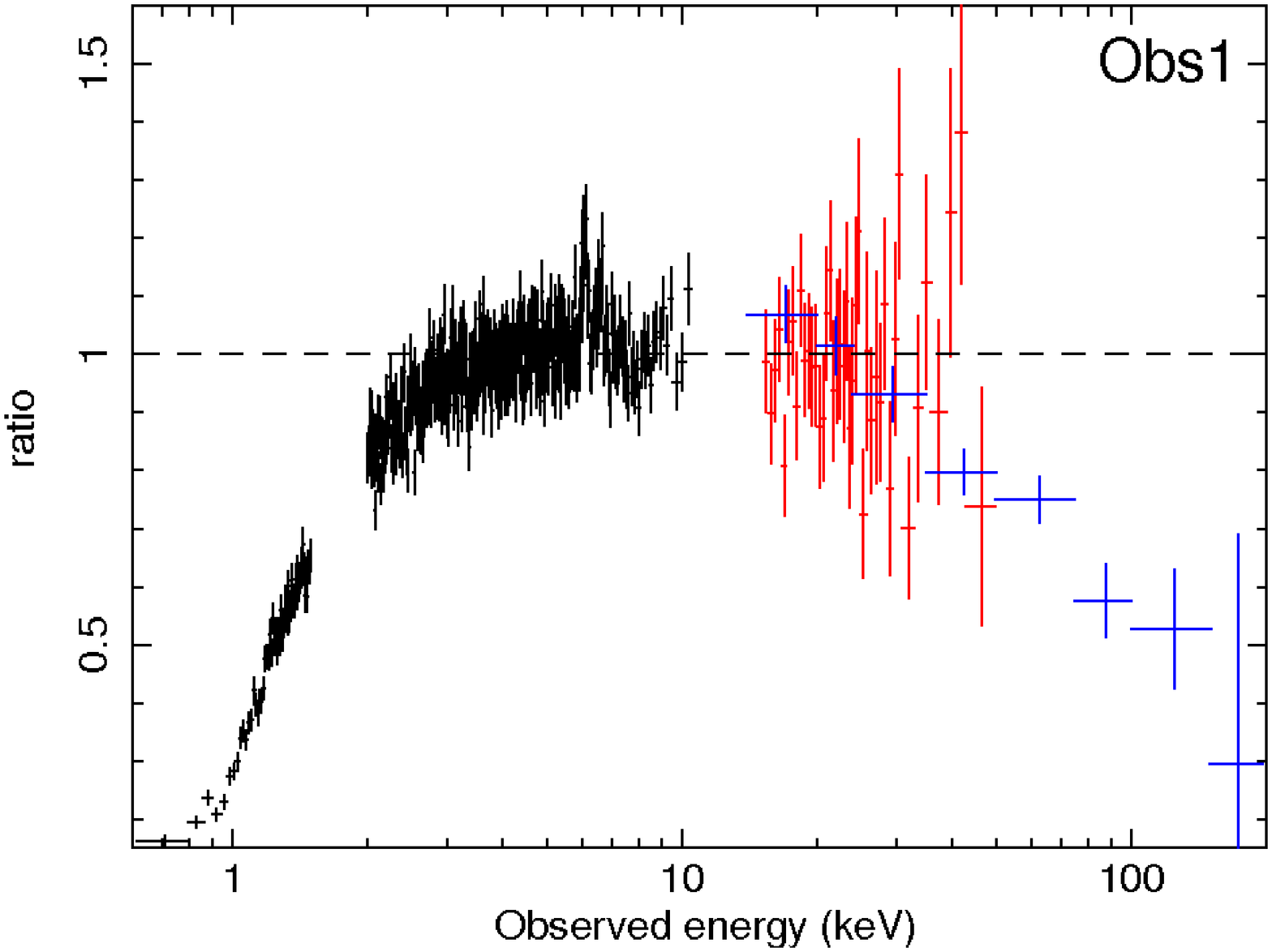}
   \includegraphics[width=4.3cm,height=5.5cm,angle=-90]{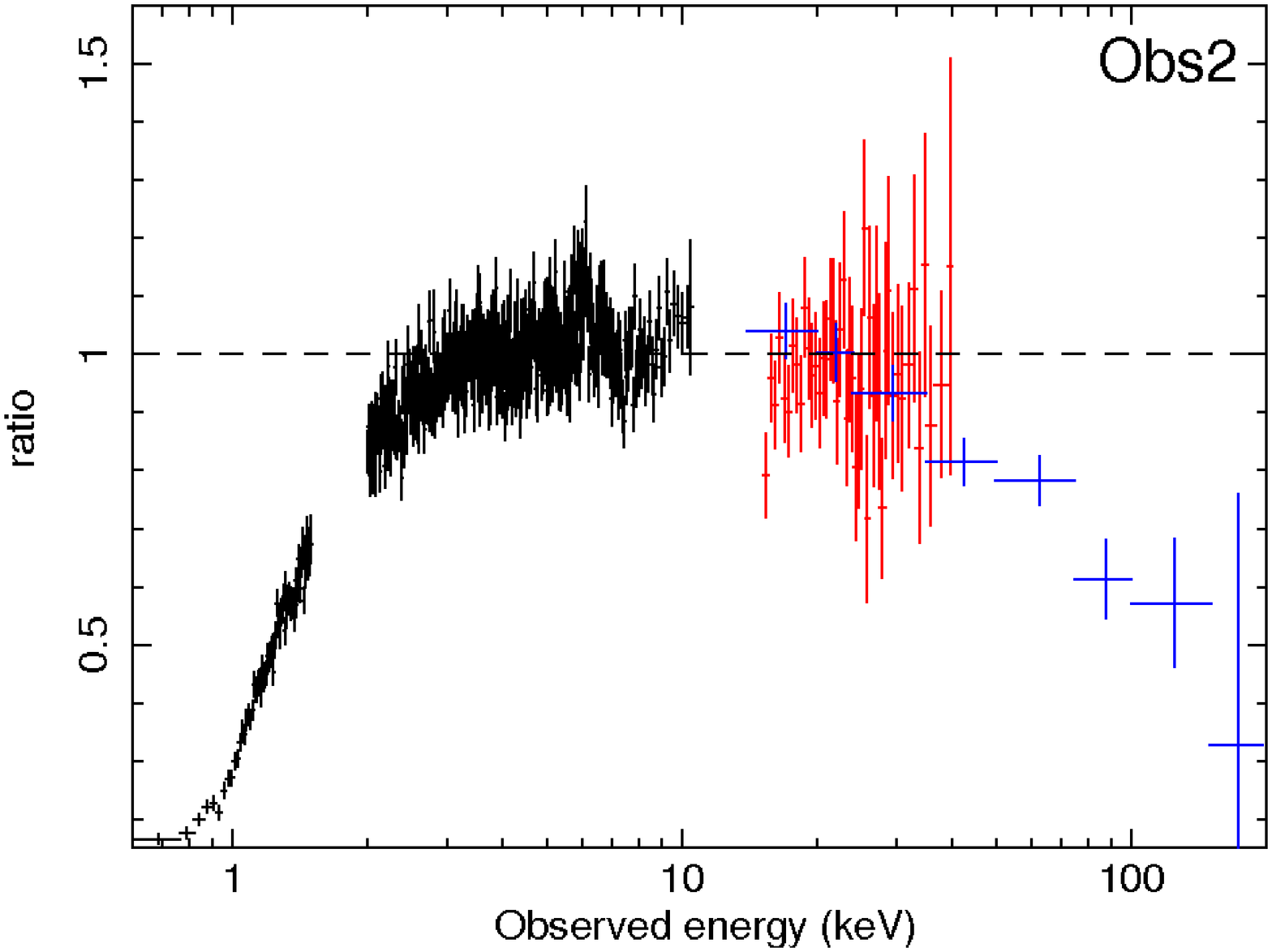}
   \includegraphics[width=4.3cm,height=5.5cm,angle=-90]{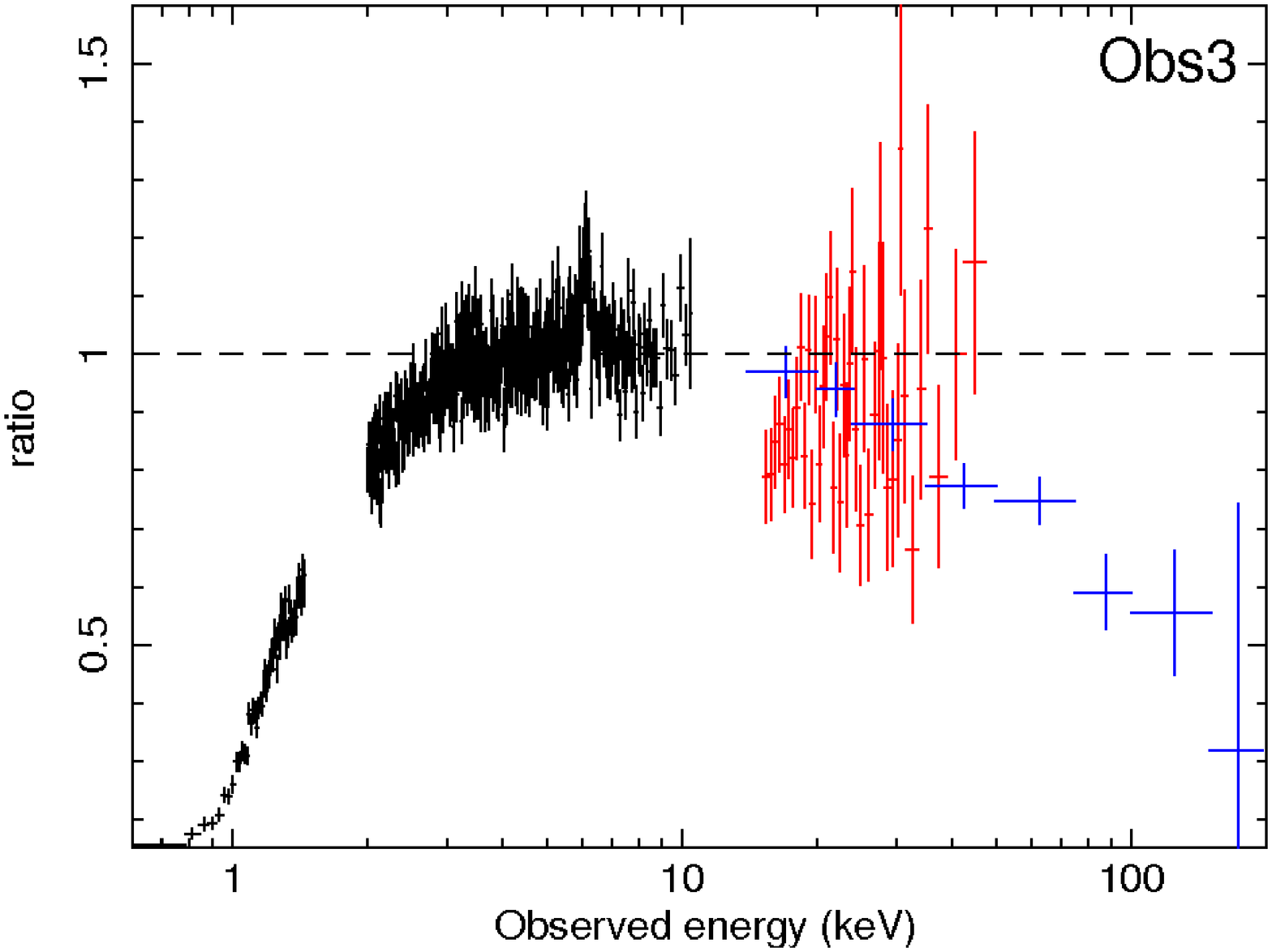}
   \caption{Ratios of the extrapolated 4--10~keV Galactic absorbed power-law continuum with respect to the \emph{Suzaku} XIS-FI (black), \emph{Suzaku} PIN (red) and \emph{Swift} BAT (blue) spectra of the three observations of 3C~111. Note that the \emph{Suzaku} XIS-FI and PIN data points have been rebinned to improve the clarity of the plots to a signal-to-noise of 20 and 5, respectively.}
    \end{figure*}

The intrinsic neutral absorption component has a column density of $N_\mathrm{H}$$\simeq$$6.5\times 10^{21}$ cm$^{-2}$. The power-law photon index becomes slightly steeper between Obs1, $\Gamma \simeq 1.67$, and Obs2 and Obs3, $\Gamma \simeq 1.74$, following also a $\sim$30\% increase in flux. The neutral reflection fraction $R$$\simeq$0.15--0.2 is equivalent between the three observations (at the 90\% level) and its weakness is consistent with the low equivalent width (EW) of the 6.4~keV Fe K$\alpha$ emission line of $\sim$40~eV (Tombesi et al.~2011b). These results are in line with the increasing number of evidences indicating that radio-loud AGNs have systematically lower reflection signatures ($R\simeq 0.1$) compared to the radio-quiet Seyferts ($R\simeq 1$; Dadina 2008), the origin of this dichotomy being still unclear (Kataoka et al.~2007; Sambruna et al.~2009, 2011; Ballo et al.~2011; Tazaki et al.~2010, 2011; Walton et al.~2013; Lohfink et al.~2013). This might suggest that the solid angle subtended by the reflector in radio-loud AGNs is low ($R\simeq \Omega/2\pi$) and/or that it is not Compton-thick. Another possibility could be that the illuminating source is outflowing with high velocity and beamed away from the disc/torus, which would possibly be identified with the base of the jet (see also \S3.3.1).

The high-energy cutoff of the primary power-law continuum is constrained to be $E_\mathrm{C}$$\sim$150--200~keV and it is consistent at the 90\% level between the observations. This is an improvement with respect to the previous spectral analysis of the 2008 \emph{Suzaku} observation presented by Ballo et al.~(2011), who could not constrain it because the source was caught with a flux 3 times lower than the present. This is consistent with what expected from thermal Comptonization models of the X-ray continuum in AGNs (e.g., Haardt \& Maraschi 1991) and it is in agreement with observations of other BLRGs and local Seyferts (Dadina 2008; Sambruna et al.~2011). Therefore, this supports other evidences that BLRGs show accretion disc/corona emission dominated spectra from optical-UV up to hard X-rays, similarly to radio-quiet Seyferts, and instead the jet contribution is dominant only at radio and $\gamma$-rays (e.g., Kataoka et al.~2007; Ballo et al.~2011; Grandi et al.~2012; de Jong, Beckmann, \& Mattana 2012).

\subsection{Ionized Fe K emission and absorption lines}

From the zoomed 5--9~keV ratios with respect to the baseline models in Fig.~3 we can observe an emission line in Obs1 at E$=$6.88~keV and an absorption line at E$=$7.75~keV in Obs2, respectively. We refer to Tombesi et al.~(2011b) for a detailed discussion of the XIS background and XIS cross-checks. The features are initially modeled as narrow Gaussian lines and their best-fit parameters are reported in Table~2. The emission line in Obs1 has a centroid energy of E$=$6.88$\pm$0.04~keV and EW$=$25$\pm$6~eV. Instead, the absorption line in Obs2 is found at E$=$7.75$\pm$0.04~keV, with EW$=$$-20$$\pm$5~eV. Assuming the same energy of this absorption line in Obs1 and Obs3 we obtain the 90\% lower limits on the EW of $>$$-8$~eV and $>$$-17$~eV, respectively. This indicates a clear variability of the absorption line, at least between the first and second observations. The fit improvements after the inclusion these lines are $\Delta\chi^2/d.o.f. = 16.5/2$ and $\Delta\chi^2/d.o.f. = 10.6/2$, corresponding to F-test confidence levels ($P_F$) of $P_F$$=$99.99\% and $P_F$$=$99.5\%, respectively. 

The detection of emission and absorption lines in these \emph{Suzaku} observations was already reported in the E$=$4--10~keV spectral analysis of Tombesi et al.~(2011b), with parameters consistent within 1$\sigma$. The detection level was a bit higher, being $\Delta\chi^2 = 25.6$ ($P_\mathrm{F}$$>$99.99\%) and $\Delta\chi^2 = 15.3 $ ($P_\mathrm{F}$$=$99.99\%), respectively. This can be ascribed to the fact that the fit in the restricted Fe K band provided a slightly better characterization of the local spectrum. However, it is important to note that there are some caveats when using the F-test method alone to study spectral lines, as it can slightly overestimate their detection significance, especially when searched over a relatively wide energy range (e.g., Protassov et al.~2002). These issues can be overcome performing extensive Monte Carlo spectral simulations (e.g., Tombesi et al.~2010b). Accordingly, Tombesi et al.~(2011b) calculated the random probability to find spurious emission/absorption features in the wide energy intervals E$=$6.5--7.5~keV and E$=$7--10~keV through 1000 Monte Carlo spectral simulations. The resultant confidence levels are $>$99.9\% and 99.8\%, respectively.

  \begin{figure*}
  \centering
   \includegraphics[width=5.7cm,height=4.7cm,angle=0]{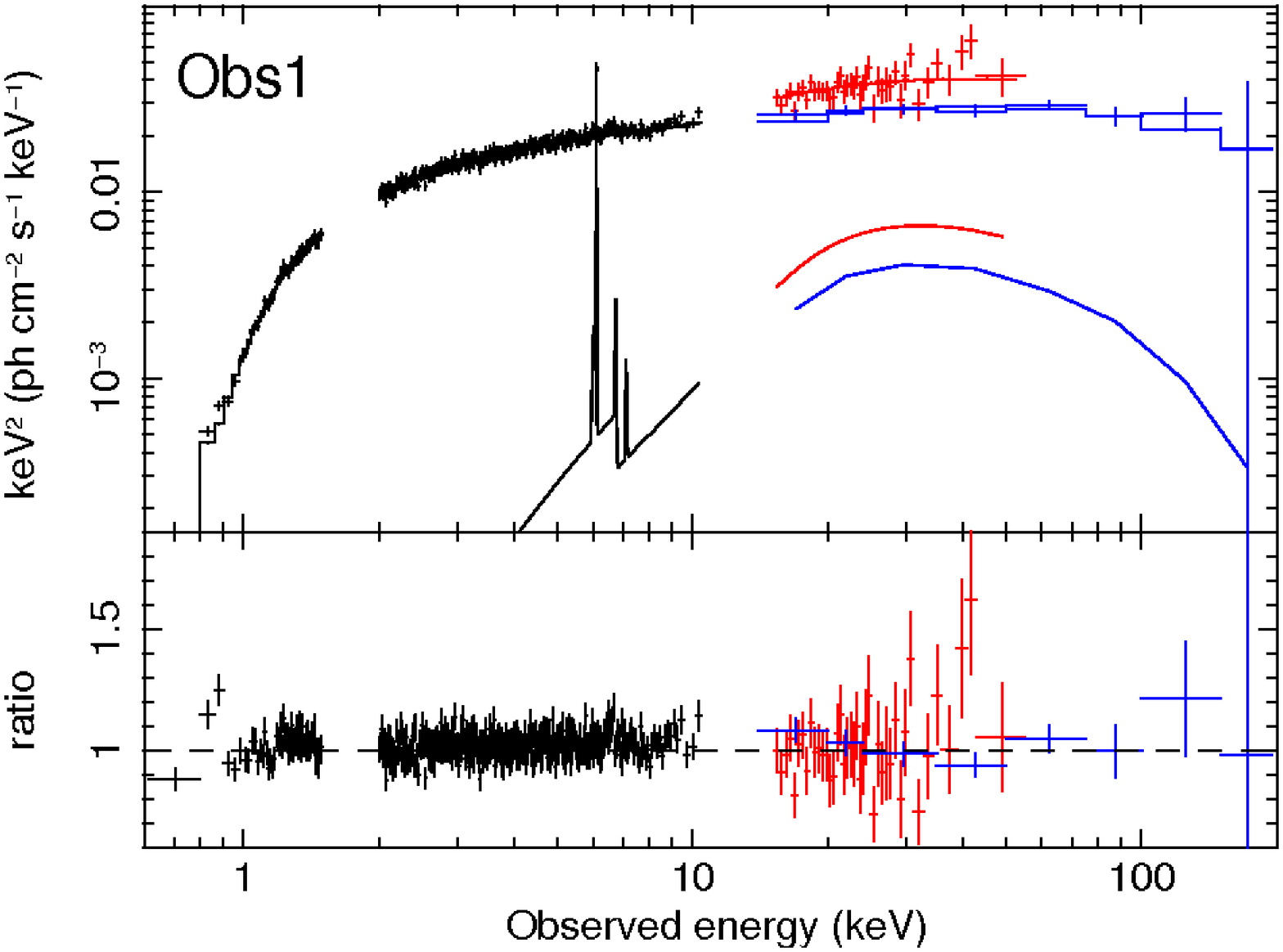}
   \includegraphics[width=5.7cm,height=4.7cm,angle=0]{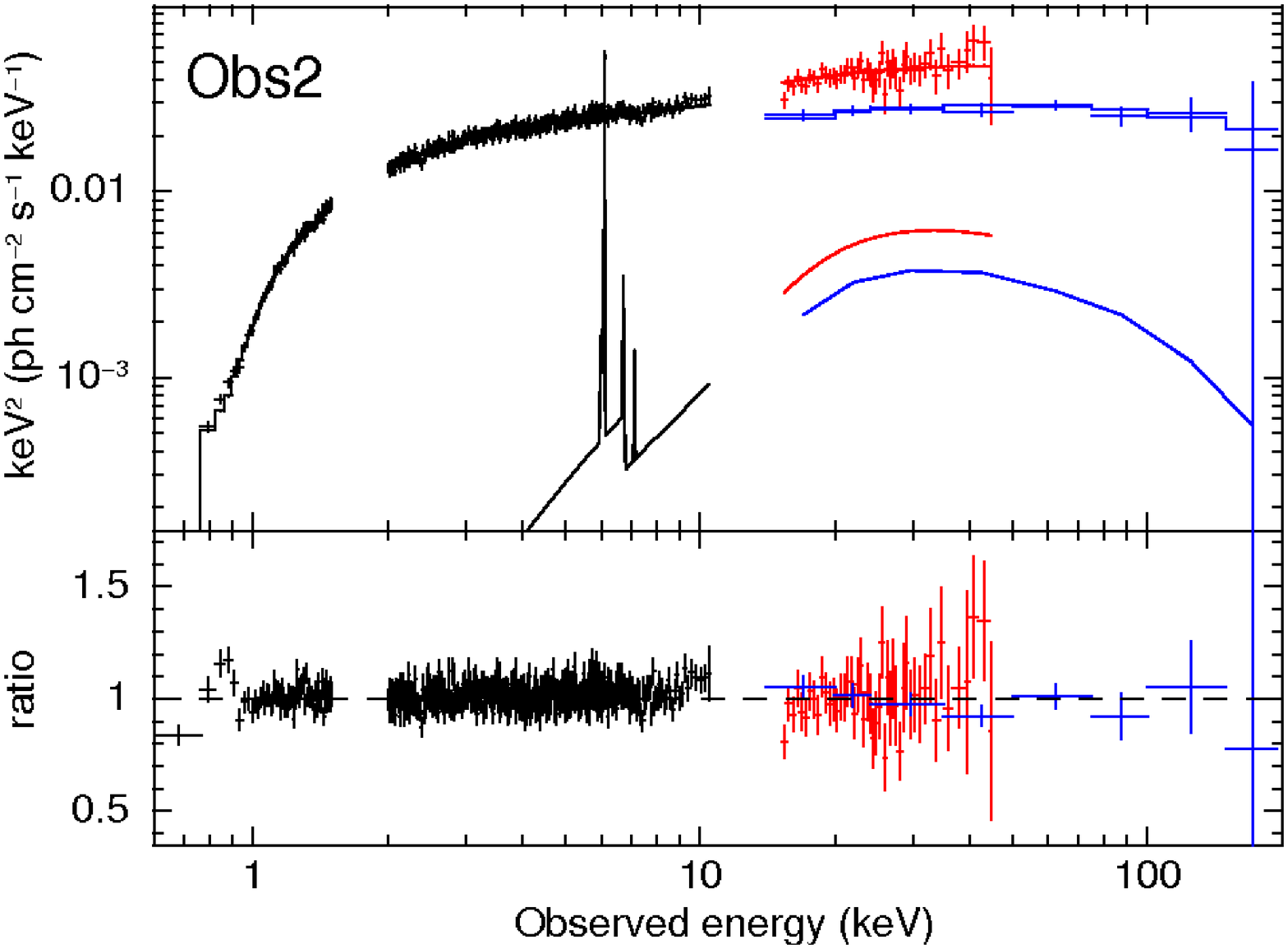}
   \includegraphics[width=5.7cm,height=4.7cm,angle=0]{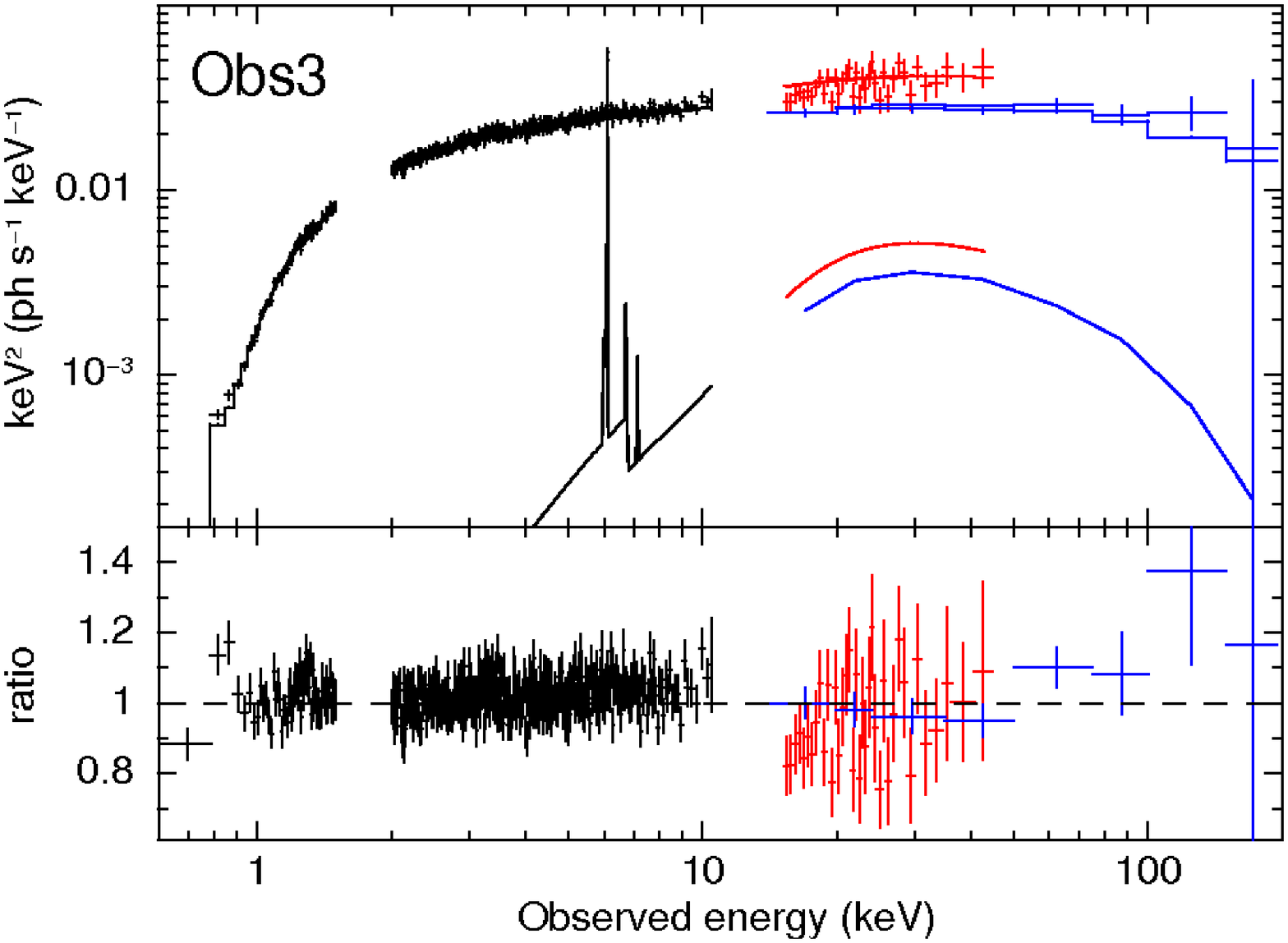}
   \caption{The upper panels show the combined \emph{Suzaku} XIS-FI (black), \emph{Suzaku} PIN (red) and \emph{Swift} BAT (blue) unfolded spectra and the relative baseline models for the three observations of 3C~111. The lower panels show the ratio of the spectra against the best-fit baseline models. Note that the \emph{Suzaku} XIS-FI and PIN data points have been rebinned to improve the clarity of the plots to a signal-to-noise of 20 and 5, respectively.}
    \end{figure*}

More recently, the detection of these lines was confirmed also by Gofford et al. (2013), who performed a systematic \emph{Suzaku} broad-band spectral analysis of a large sample of 51 AGNs. The line parameters are still consistent within 1$\sigma$. The fit improvements are also consistent with those reported here, $\Delta\chi^2 = 13.0$ ($P_\mathrm{F}$$>$99.99\%) and $\Delta\chi^2 = 11.1$ ($P_\mathrm{F}$$=$99.6\%) for the emission and absorption lines, respectively. Gofford et al.~(2013) also performed an additional extensive Monte Carlo analysis of the significance of the absorption line, obtaining a probability of 97.8\%. Although not very high, this is larger than the Monte Carlo probability threshold of 95\% used by both Tombesi et al.~(2010b) and Gofford et al.~(2013) for the blue-shifted Fe XXV/XXVI absorption lines. Therefore, even if there are some discrepancies due to the different energy bands considered and the models employed, these results all support the reality of these spectral features.

The energies of the observed lines are not directly consistent with any known atomic transition, however they are close to Fe XXVI Ly$\alpha$ at 6.97~keV (Kallman et al.~2004). In the next sections we apply more complex models to derive a better self-consistent interpretation of these lines and discuss their possible physical origin(s). Throughout the paper we use the negative sign to indicate redshift, if not otherwise specified.

\subsection{Modelling the ionized emission in Obs1}

\subsubsection{Ionized disc reflection}

In Tombesi et al.~(2011b) we show the possibility to model the ionized Fe K emission line in Obs1 as the blue peak of a relativistic disc line with $r_\mathrm{in}$$\sim$20--100~$r_g$. We test again this identification here using the relativistic line profile \emph{relline} in XSPEC (Dauser et al.~2010). Given the limited signal-to-noise of the data, the black hole spin, emissivity profile and outer radius cannot be constrained and were fixed to the typical values of $a$$=$0.99 for rapidly spinning black holes (e.g., Reynolds 2013; Walton et al.~2013), $\beta$$=$$-$3 and $r_\mathrm{out}$$=$400~$r_g$ ($r_g \equiv GM_\mathrm{BH}/c^2$, which corresponds to $r_g$$\simeq$$1.8\times 10^{13}$~cm for the SMBH in 3C~111), respectively. We assume an inclination angle consistent with that of the jet of $\simeq$18$^{\circ}$ (Jorstad et al.~2005). Assuming in turn the energy of Fe XXV He$\alpha$ at 6.7~keV or Fe XXVI Ly$\alpha$ at 6.97~keV we find a significant ($\Delta\chi^2\simeq 10$) improvement in the latter case, indicating that the data favor the Fe~XXVI interpretation. If we leave the energy free, we obtain a best-fit with E$=$6.92$\pm$0.04~keV, $I$$=$$(1.4\pm0.3)\times 10^{-5}$ ph~s$^{-1}$~cm$^{-2}$ and EW$=$28$\pm$6~eV. This provides a fit improvement with respect to the baseline model of $\chi^2/d.o.f.$$=$15.7/3 ($P_\mathrm{F}$$\simeq$99.9\%). However, we can only place a 90\% lower limit to the inner radius of the disc reflecting surface of $r_\mathrm{in}$$\ge$50~$r_g$. This is consistent with previous claims in other radio galaxies, such as 3C~120 (Kataoka et al.~2007; Lohfink et al.~2013), 3C~390.3 (Sambruna et al.~2009), 4C$+$74.26 (Larsson et al.~2008) and 3C~382 (Sambruna et al.~2011). 

We tried also a fit using the lamp post geometry, \emph{relline\_lp} in XSPEC (Dauser et al.~2013), in which the accretion disc is illuminated by a compact source located ontop the black hole, possibly identifiable with the base of a jet. Very broad lines are produced only for compact irradiating sources situated very close to the black hole, at heights $h$$\la$10~$r_g$ (e.g., Wilkins \& Fabian 2012, 2013). We assume again $a$$=$0.99, an inclination of 18$^{\circ}$ and outer radius of $r_\mathrm{out}$$=$400~$r_g$. We fix the inner radius, $r_\mathrm{in}$, to the innermost stable circuar orbit (ISCO), which corresponds to $r_\mathrm{ISCO}$$\simeq$1.23~$r_g$ for $a$$\simeq$0.99. This does not affect the final results, given that the fit is insensitive to this parameter. Instead, the height of the X-ray source, $h$, is free to vary.
Assuming in turn the energy of Fe XXV He$\alpha$ at 6.7~keV or Fe XXVI Ly$\alpha$ at 6.97~keV we find again a significant ($\Delta\chi^2\simeq 13$) improvement in the latter case. Leaving the energy free, we obtain E$=$6.94$^{+0.05}_{-0.04}$~keV, I$=$$(1.3^{+0.4}_{-0.2})\times 10^{-5}$ ph~s$^{-1}$~cm$^{-2}$ and EW$=$26$\pm$6~eV. The 90\% lower limit on the height of the illuminating source is found at $h$$\ga$80~$r_g$. The fit improvement is equivalent to the prevous one. 
Given the uncertainty in the inclination derived from the jet, $18^{\circ}\pm5^{\circ}$ (Jorstad et al.~2005) or 10$^{\circ} < i < 26^{\circ}$ (Lewis et al.~2005), we tried also to leave this parameter free. We obtain an equivalent fit, with  $\Delta\chi^2/d.o.f. = 16.4/4$. The parameters of the line are E$=$6.97$\pm$0.06~keV, $I$$=$$(1.3\pm0.3)\times 10^{-5}$  ph~s$^{-1}$~cm$^{-2}$ and EW$=$27$\pm$6~eV. The 90\% upper and lower limits on the inclination and source height are $i$$\le$17$^{\circ}$ and $h$$\ge$30~$r_g$, respectively. The inclination is consistent with the jet estimate. 

It is important to note that from these fits it is unclear whether the inner disc is truncated or not, as equivalent fits are obtained with $r_\mathrm{in}$$\ge$50~$r_\mathrm{g}$ or $r_\mathrm{in}$ extending down to the ISCO for a maximally spinning black hole in the lamp-post scenario. In particular, the fact that the disc should not be strongly truncated is supported by studies of Galactic black hole binaries with comparable Eddington ratios (e.g., Reis et al.~2010, 2011, 2012; Walton et al.~2012; Reynolds \& Miller 2013).

\begin{table*}
\centering
\begin{minipage}{90mm}
\caption{Best-fit parameters of the baseline broad-band models.}
\begin{tabular}{lccc}
\hline\hline
 Model & Obs1 & Obs2 & Obs3 \\
\hline\hline
\multicolumn{4}{l}{Galactic absorption ($10^{22}$~cm$^{-2}$)} \\
\hline
$N_H$  & 0.3 & 0.3  & 0.3\\
\hline\hline
\multicolumn{4}{l}{Neutral absorption ($10^{22}$~cm$^{-2}$)} \\
\hline
$N_H$  & 0.64$\pm$0.01 & 0.65$\pm$0.01 & 0.68$\pm$0.01\\
\hline\hline
\multicolumn{4}{l}{Pexmon} \\
\hline
$\Gamma$ & 1.67$\pm$0.01 & 1.73$\pm$0.01 & 1.74$\pm$0.01 \\
$R$ & 0.20$\pm$0.02 & 0.14$\pm$0.02 & 0.15$\pm$0.02\\
$E_C$ (keV) & 154$^{+24}_{-19}$ & 219$^{+47}_{-34}$ & 135$^{+22}_{-17}$\\ 
\hline\hline
\multicolumn{4}{l}{Best-fit baseline models}\\
\hline
$\chi^2/d.o.f.$ & 2195.6/2149 & 2318.6/2244 & 2241.5/2180\\
\hline\hline
\multicolumn{4}{l}{Gaussian emission/absorption lines} \\
\hline
$E$ (keV) & 6.88$\pm$0.04 & 7.75$\pm$0.04 & \\
$\sigma$ (eV) & $<$150$^a$ & $<$300$^a$ & \\ 
$I$ ($10^{-5}$ph~s$^{-1}$cm$^{-2}$) & 1.3$\pm$0.3 & $-$1.1$\pm$0.3 & \\
$EW$ (eV) & 25$\pm$6 & $-$20$\pm$5 & \\
\hline\hline
\multicolumn{4}{l}{Best-fit including emission/absorption lines}\\
\hline
$\chi^2/d.o.f.$ & 2179.1/2147 & 2308.0/2242 & \\
\hline\hline
\multicolumn{4}{l}{Flux/Luminosity$^b$ ($10^{-11}$erg~s$^{-1}$cm$^{-2}$/$10^{44}$erg~s$^{-1}$)} \\
\hline
0.5--2~keV  & 2.6/1.4  & 3.7/2.0  & 3.6/1.9\\
2--10~keV   & 4.7/2.6  & 6.2/3.4  & 6.0/3.3\\
15--200~keV & 11.0/6.0 & 12.0/6.7 & 12.0/6.5\\
\hline
\end{tabular}
{\em Note.} $^a$ 90\% upper limit. $^b$ Absorption corrected continuum flux and luminosity. 
\end{minipage}
\end{table*}

However, in order to derive a more physically self-consistent characterization of the possible Fe K relativistic line, we need to perform a fit using detailed broad-band ionized disc reflection models, such as \emph{reflionx} (Ross \& Fabian 2005) and the more recent \emph{xillver} model (Garc{\'{\i}}a \& Kallman 2010; Garc{\'{\i}}a et al.~2013). Here, we show fits with the latter model, but equivalent results are obtained with both. We use the standard definition of the ionization parameter $\xi$$=$$L_\mathrm{ion}/n r^2$ erg~s$^{-1}$~cm (Tarter et al. 1969) where $L_\mathrm{ion}$ is the ionizing luminosity between 1 Ryd and 1000 Ryd (1 Ryd $=$ 13.6 eV), $n$ is the number density of the material, and $r$ is the distance of the gas from the central source. 

During the fits, the slope of the ionizing continuum in \emph{xillver} is linked to the source power-law continuum. First, we do not apply any relativistic blurring to this reflected emission. Assuming standard Solar abundances, we obtain a marginal fit improvement from the addition of this component, $\chi^2/d.o.f.$$=$2191.5/2147. The ionization parameter is estimated to be log$\xi$$\ga$3.9 erg~s$^{-1}$~cm at the 90\% level. Leaving the iron abundance and redshift free to vary we obtain a slight improvement of $\Delta\chi^2/d.o.f. \simeq 6.1/4$ with respect to the baseline model fit. The best-fit parameters now indicate an iron abundance $>$2 times Solar (at the 90\% level) and log$\xi$$=$$3.49^{+0.43}_{-0.32}$ erg~s$^{-1}$~cm, indicating Fe XXVI emission (e.g., Garc{\'{\i}}a \& Kallman 2010). However, the inclusion of this reflection component is only marginally required at the $\simeq$90\% level. This is consistent with the fact that the ratios with respect to the baseline model including a weak ($R\simeq 0.2$) neutral reflection component in Fig.~2 do not indicate any significant additional excess at E$\simeq$10--100~keV, where the reflection hump would show up. The best-fit broadband model is shown in Fig.~4 (panel a) and the relative parameters without relativistic blurring are reported in Table~3. The inclusion of relativistic blurring (e.g., Dauser et al.~2010) does not provide any fit improvements, giving an inner radius consistent with the outer radius of 400~$r_\mathrm{g}$. As we can see from Fig.~4 (panel a), the Fe K emission line from the Compton-thick reflection table is already intrinsically quite broad due to the large electron scattering depth at such high values of ionization and iron abundance.

Therefore, even if the line profile can be possibly modeled as an ionized disc line, a good fit can not be found with standard, broad-band ionized disc reflection models, even without relativistic blurring applied. Some modelling assumptions might complicate the fit. For instance, the \emph{Xillver} model considers only the case of a Compton-thick, constant density disc atmosphere at a single ionization state. In the next sections we explore alternative possibilities.

  \begin{figure*}
  \centering
   \includegraphics[width=4.7cm,height=5.7cm,angle=-90]{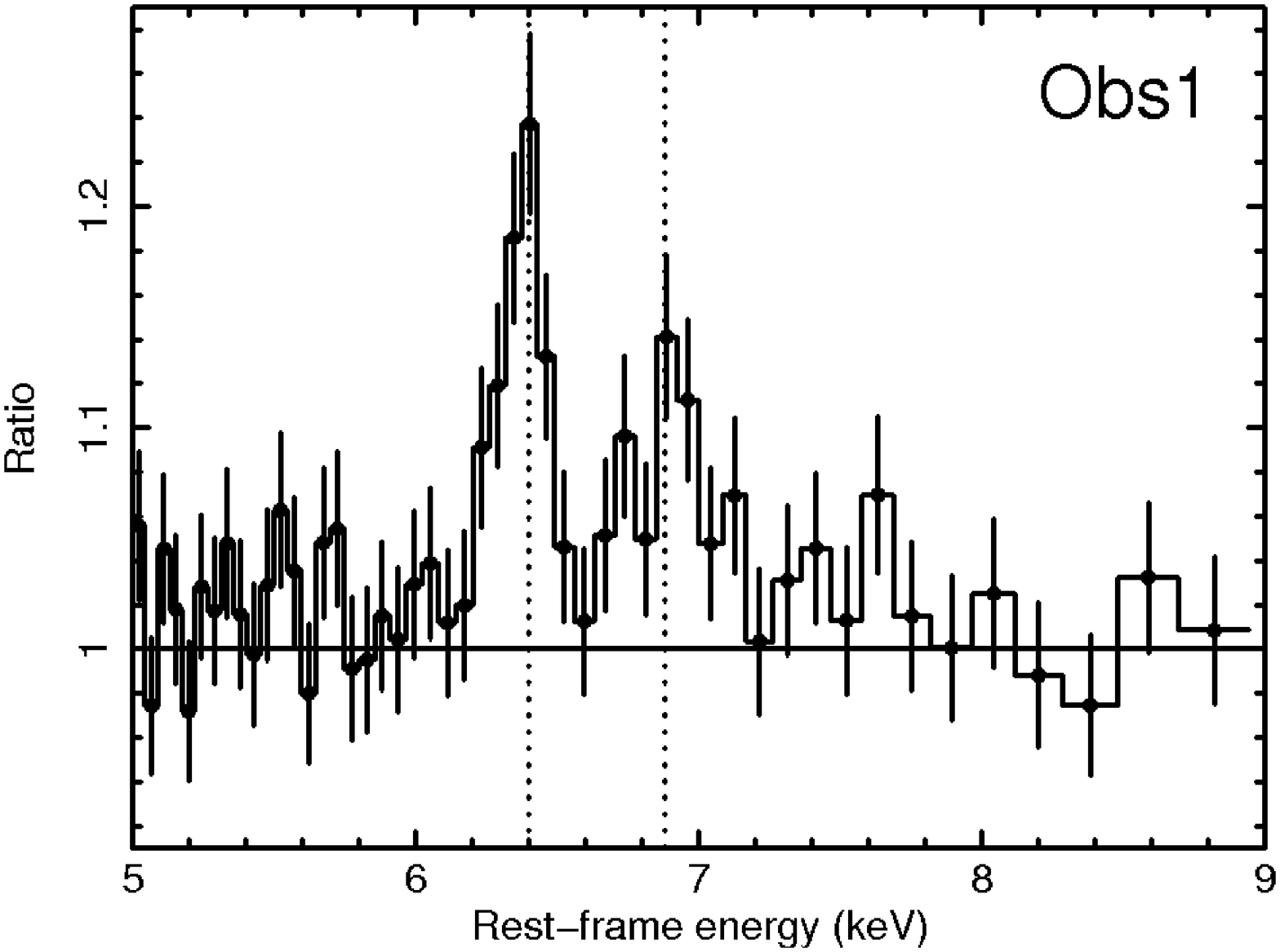}
   \includegraphics[width=4.7cm,height=5.7cm,angle=-90]{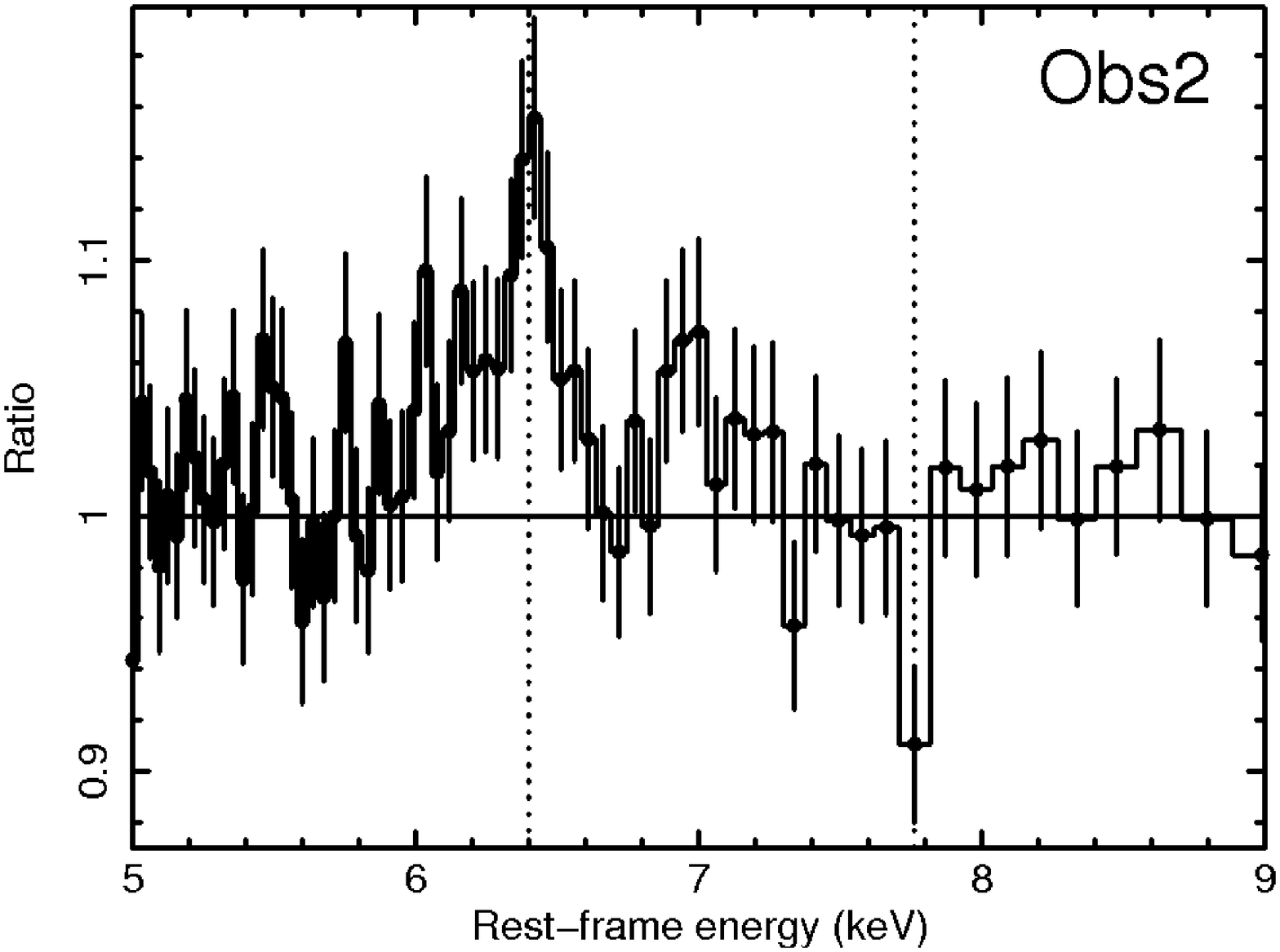}
   \includegraphics[width=4.7cm,height=5.7cm,angle=-90]{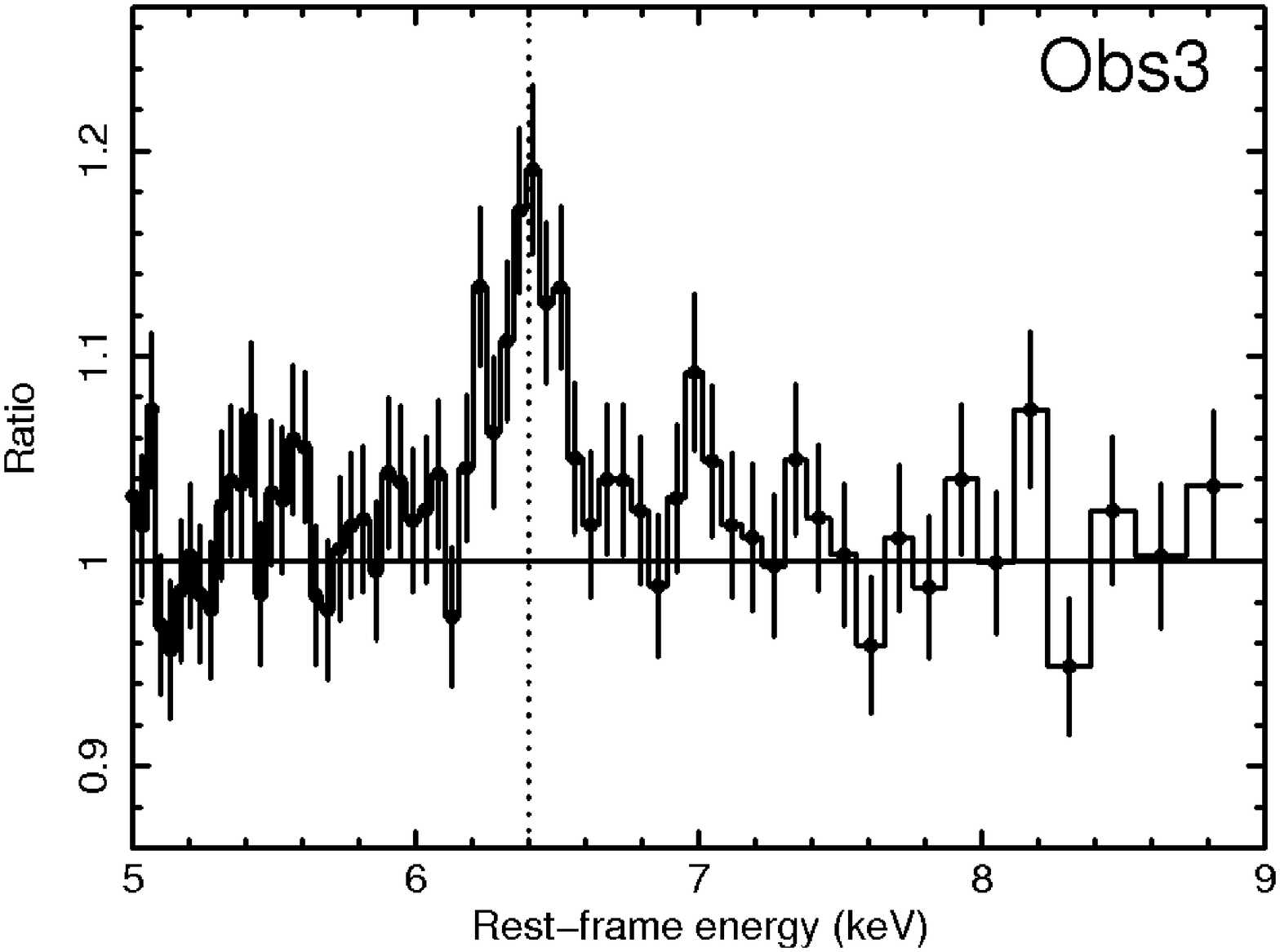}
   \caption{Ratios of the broad-band \emph{Suzaku} XIS-FI spectra with respect an absorbed power-law continuum and cold reflection (\emph{pexrav} in XSPEC) model zoomed in the rest-frame E$=$5--9~keV band. The vertical dotted lines refer to the detected lines. Note that the data points have been significantly rebinned to a signal-to-noise of 30 to improve the clarity of the plots.}
    \end{figure*}

\subsubsection{Hot collisional plasma}

We investigate the alternative origin of the emission line from a high temperature, collisionally ionized plasma (e.g., Lobban et al.~2010; Emmanoulopoulos et al.~2012). Equivalent results are obtained using the publicly available models \emph{apec} (Smith et al.~2001) and \emph{mekal} (Mewe et al.~1985) in XSPEC. The addition of this component provides a fit improvement of $\Delta\chi^2/d.o.f. = 13.5/3$ ($P_\mathrm{F}$$=$99.6\%). The temperature required to produce such line is very high, $kT=22.0^{+6.1}_{-3.2}$~keV, indicating again an interpretation as Fe XXVI. If let free, the elemental abundance  is consistent with the standard Solar value, $A$$=$$0.83^{+0.55}_{-0.34}$, and the redshift with $z=0.058\pm0.006$. The best-fit broadband model is shown in Fig.~4 (panel b) and the relative parameters are reported in Table~3. Therefore, a simple hot plasma component provides a fit that is statistically better than the one previously obtained using an ionized disc reflection model. However, it is not clear what would be the physical origin of this very hot material (see \S4.2.2).

\subsubsection{Photo-ionized emission}

The observed emission line could alternatively originate from a Compton-thin, photo-ionized plasma (e.g., Lobban et al.~2010; Lohfink et al.~2012). In type 1 sources these features are more diluted due to the higher continuum flux than in type 2s, which are partially obscured (e.g., Bianchi et al.~2003). In particular, Bianchi \& Matt (2002) showed that an equivalent width of $\sim$15--20~eV for H-like Fe could be expected from material with $N_H \sim 10^{23}$~cm$^{-2}$ and log$\xi \sim$4~erg~s$^{-1}$~cm. 

Therefore, we test this scenario adding a photo-ionized emission component to the baseline model of Obs1 using the XSTAR code v.~2.2.1bk (Kallman \& Bautista 2001). This takes into account both the transmitted and reflected emission from a Compton-thin plasma. We assume a typical power-law continuum with $\Gamma$$=$2, E$_c$$\sim$100~keV and standard Solar abundances (Asplund et al.~2009). We test two turbulent velocities of 100~km/s and 1,000~km/s, the former providing a better fit ($\Delta\chi^2$$>$5). The free parameters of this photo-ionization component are the ionization parameter, the column density and the normalization. However, we note that there is a degeneracy between the latter two. 

This is easy to overcome because the normalization depends on a few other known quantities. For a uniform, spherical symmetric, Compton-thin shell, the normalization of the photo-ionized emission component is defined within the XSTAR code by\footnote{http://heasarc.gsfc.nasa.gov/xstar/docs/html/node94.html} $K = f L_\mathrm{ion}/D^2$, where $f$$=\Omega/4\pi$ is the covering fraction of the material, $L_\mathrm{ion}$ is the ionizing luminosity in units $10^{38}$~erg~s$^{-1}$ from 1 to 1,000 Ryd and $D$ is the distance of the observer to the source in kpc. The emitter normalization and geometry are not uniquely determined. However, assuming a covering fraction $f = 1$ we can then estimate a lower limit in the column density. Substituting the appropriate values of distance $D \simeq 2\times 10^5$~kpc derived applying the Hubble law to the cosmological redshift of 3C~111 and its ionizing luminosity $L_\mathrm{ion} \simeq 6.1\times 10^{6} (\times 10^{38})$~erg~s$^{-1}$, we obtain $K \simeq 1.5\times 10^{-4}$. 

We obtain a fit improvement of $\Delta\chi^2/d.o.f. = 14.5/3 $ ($P_\mathrm{F}$$=$99.7\%), with $N_\mathrm{H} > 3\times 10^{23}$~cm$^{-2}$ (at 90\%) and log$\xi$$=$$4.52^{+0.10}_{-0.16}$~erg~s$^{-1}$~cm. The favored line identification is again Fe XXVI Ly$\alpha$ and the redshift is $z=0.059\pm0.006$, consistent with the source cosmological one at 90\%. The best-fit broadband model is shown in panel c of Fig.~4 and the relative parameters are reported in Table~3. The fit with this photo-ionized emission component provides the lowest $\chi^2/d.o.f.$ among the three cases reported in Table~3.

\subsection{Modelling the ionized absorption in Obs2}

Following Tombesi et al.~(2011b), we model the highly ionized Fe K absorber in Obs2 with the photoionization code XSTAR assuming a standard $\Gamma \simeq 2$ power-law ionizing continuum with cut-off at $\sim$100~keV and standard Solar abundances (Asplund et al.~2009). Due to the limited energy resolution of the XIS, the absorption line is not resolved and we can place only a loose 90\% upper limit of $\sigma_\mathrm{v} \le$10,000~km~s$^{-1}$ . Therefore, we performed fits using three turbulent velocities of 1000, 3000 and 5000~km~s$^{-1}$, respectively. The best-fit solution is provided by the table with turbulent velocity of 3000~km~s$^{-1}$ ($\chi^2/d.o.f. = 11.6/3$, $P_f$$=$99\%), with $v_\mathrm{out}$$=$$0.104\pm0.006$c, log$\xi$$=$$4.47^{+0.76}_{-0.04}$ erg~s$^{-1}$~cm and $N_\mathrm{H}$$=$$(5.3^{+1.8}_{-1.3})\times 10^{22}$ cm$^{-2}$. This indicates an identification of the absorption line with blue-shifted Fe~XXVI Ly$\alpha$, possibly from a mildly-relativistic UFO. These parameters are consistent with those derived by Tombesi et al.~(2011b) and Gofford et al.~(2013) for the same dataset.

In Fig.~5 with show the two-dimensional contour plots of log$\xi$ vs. $N_\mathrm{H}$.  The absolute minimum in the $\chi^2$ distribution is relatively well constrained within the 1 sigma contours. However, if we consider the 90\% level, the errors expand significantly to log$\xi$$=$$4.47^{+1.36}_{-0.07}$ erg~s$^{-1}$~cm and $N_\mathrm{H}$$=$$(5.3^{+3.2}_{-2.0})\times 10^{22}$ cm$^{-2}$. However, from Fig.~5 we note that other solutions for $N_\mathrm{H} > 10^{23}$~cm$^{-2}$ may possibly be allowed. Therefore, we can conservatively say that $N_\mathrm{H} > 3\times10^{22}$~cm$^{-2}$ at the 90\% level.

\section{Discussion}

For a photo-ionized gas observed in both emission and absorption, the photo-ionization and recombination time-scales are very rapid, much shorter than the 7~days time interval between the observations (e.g., Krolik \& Kallman 1987; Shull \& Van Steenberg 1982; Garc{\'{\i}}a, Bautista, \& Kallman 2012). From the observed $\simeq$7~days spectral variability and the light-crossing time argument, we can derive a typical length-scale of $r \la c\Delta t \la 1.8\times 10^{16}$~cm ($\la 10^3$~$r_g$) for both emission and absorption (e.g., Tombesi et al.~2011b). 

\subsection{The ultra-fast outflow in Obs2}

The modelling of the blue-shifted Fe~XXVI absorption line in Obs2 with the photo-ionization code XSTAR clearly indicates the presence of a UFO with mildly-relativistic outflow velocity of $v_\mathrm{out}$$\simeq$0.1c. The lower and upper limits of the distance, mass outflow rate and kinetic power of this UFO can be estimated following the method outlined in Tombesi et al.~(2012a; 2013). In particular, an upper and lower limit on the location can be derived from $r_{\mathrm{max}} \equiv L_{\mathrm{ion}}/\xi N_\mathrm{H}$ and $r_{\mathrm{min}} \equiv 2 G M_{\mathrm{BH}}/ v_{\mathrm{out}}^{2}$ (e.g. Crenshaw \& Kraemer 2012). The mass outflow rate from a biconical wind-like geometry can be expressed as $\dot{M}_{\mathrm{out}} \simeq \pi \mu m_\mathrm{p} N_\mathrm{H} v_\mathrm{out} r$ (Krongold et al.~2007). 

Substituting the relative values and an absorption-corrected ionizing luminosity for Obs2 extrapolated from the broad-band spectrum of $L_\mathrm{ion}$$\simeq$$8.8\times 10^{44}$~erg~s$^{-1}$, we obtain $r_\mathrm{min}$$\simeq$$3.7\times 10^{15}$~cm ($\simeq$200~$r_g$), $r_\mathrm{max}$$\simeq$$5.6\times 10^{17}$~cm ($\sim$0.2~pc) and $\dot{M}_\mathrm{out}$$\simeq$0.05--7~$M_{\odot}$~yr$^{-1}$. The mechanical power of the UFO can consequently be derived as $\dot{E}_\mathrm{K} \equiv \frac{1}{2} \dot{M}_{\mathrm{out}} v_\mathrm{out}^2$$\simeq$$1.4\times 10^{43}$--$2.1\times 10^{45}$ erg~s$^{-1}$. Considering the maximum distance from the observed spectral variability of $\sim$7~days and the light crossing time argument, we obtain $r$$\simeq$200--1000~$r_g$, $\dot{M}_\mathrm{out}$$\simeq$0.05--0.3~$M_{\odot}$~yr$^{-1}$ and $\dot{E}_\mathrm{K}$$\simeq$$1.4\times 10^{43}$--$9.0\times 10^{43}$~erg~s$^{-1}$. Alternatively, equivalent results are obtained assuming that the observed variability is due to the line-of-sight motion of an absorbing cloud.


Estimating the bolometric luminosity $L_\mathrm{bol}$$\simeq$$8.8\times 10^{45}$~erg~s$^{-1}$ from $L_\mathrm{bol} = k_\mathrm{bol} L_\mathrm{ion}$, where $k_\mathrm{bol}$$\simeq$10 (McKernan et al.~2007; Vasudevan \& Fabian 2009; Lusso et al.~2010), we find that the the mechanical power of the UFO corresponds to at least $\sim$1\% of this value. This is consistent with the fraction of $\sim$0.5--5\% required by numerical simulations of AGN feedback driven by winds/outflows (e.g., Di Matteo et al.~2005; Hopkins \& Elvis 2010; Wagner et al.~2013).

  \begin{figure}
  \centering
   \includegraphics[width=6.2cm,height=8.2cm,angle=-90]{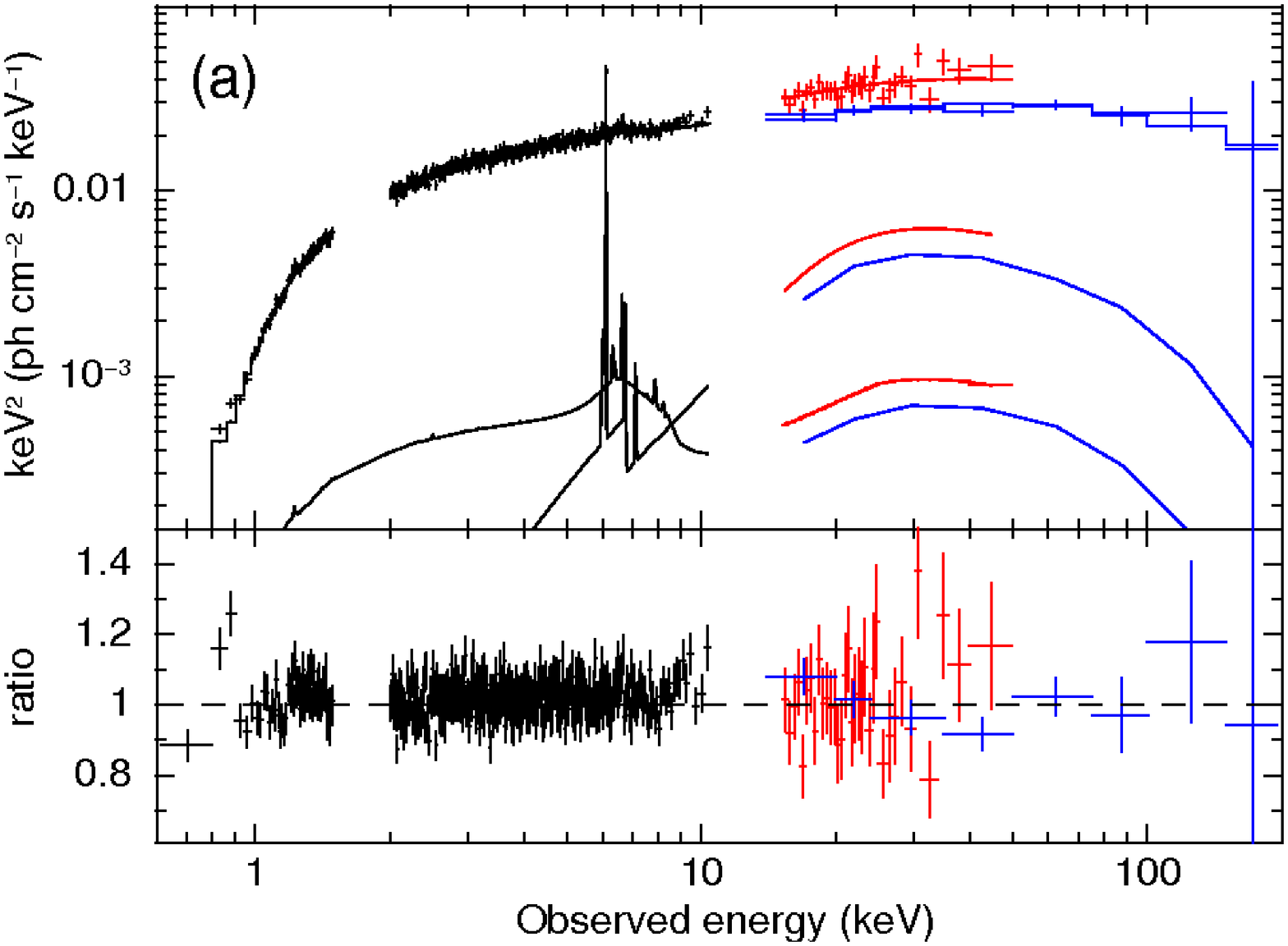}
\hspace{0.5cm}
   \includegraphics[width=6.2cm,height=8.2cm,angle=-90]{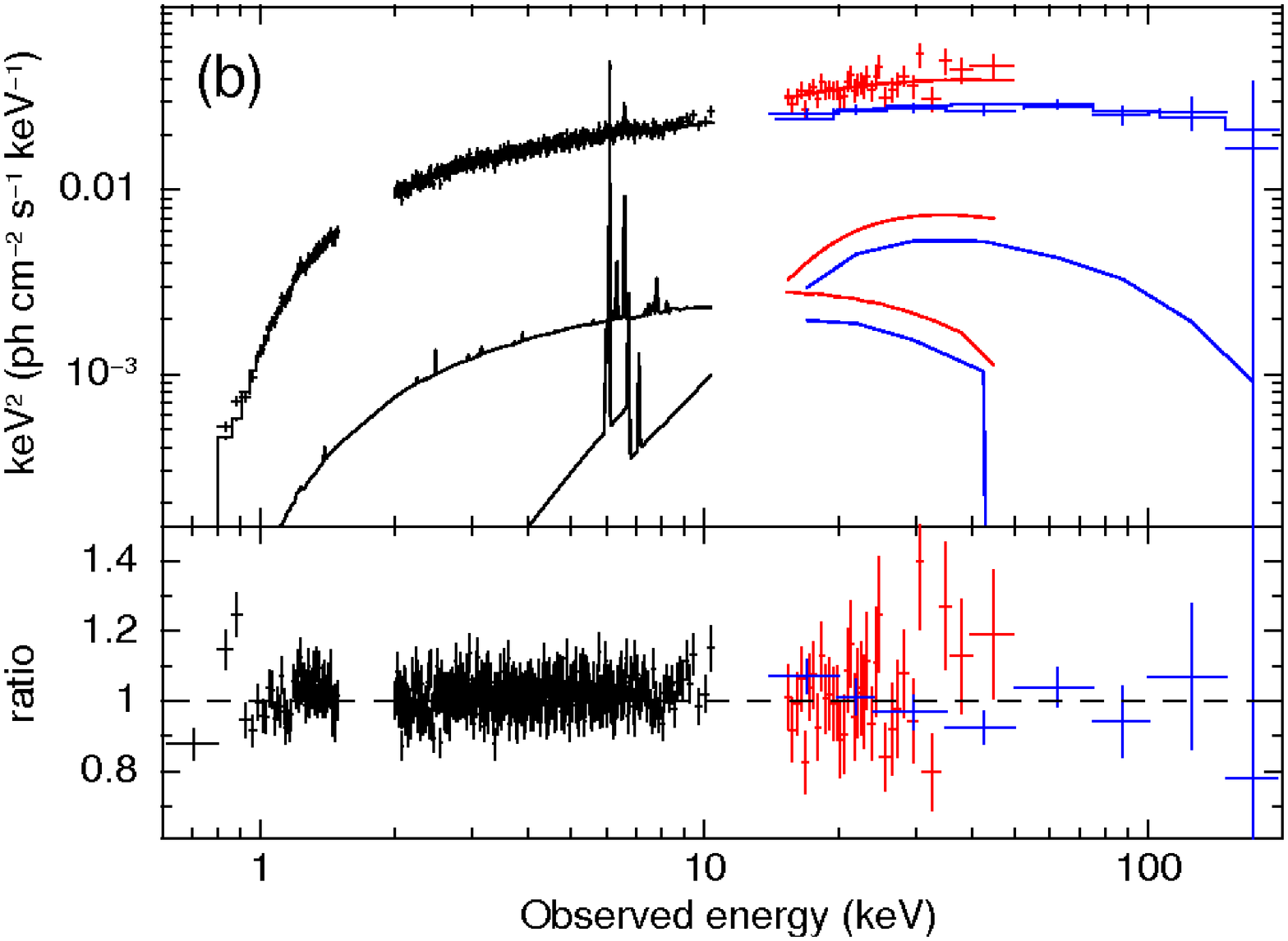}
\hspace{0.5cm}
   \includegraphics[width=6.2cm,height=8.2cm,angle=-90]{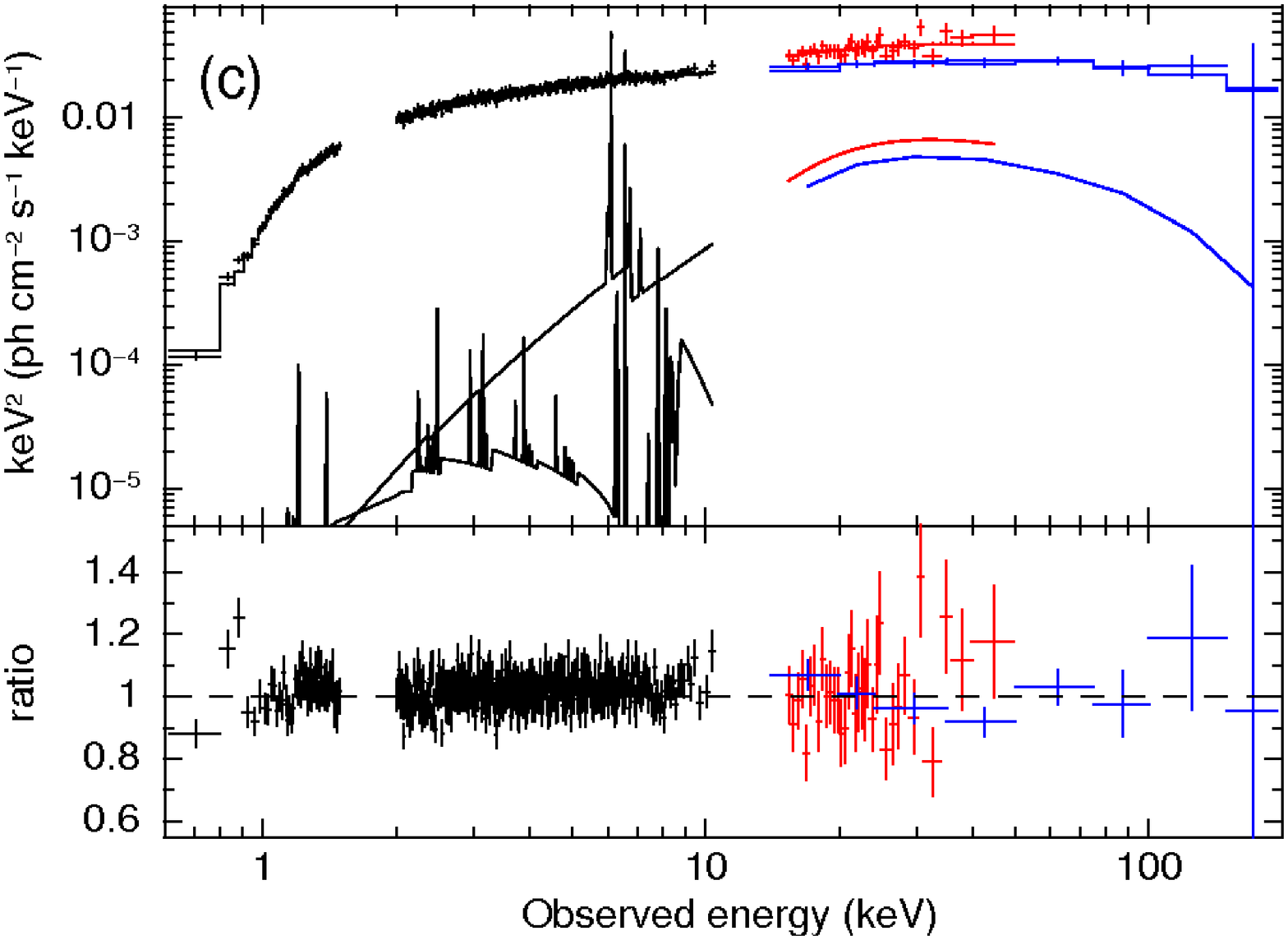}
   \caption{Combined \emph{Suzaku} XIS-FI (black), \emph{Suzaku} PIN (red) and \emph{Swift} BAT (blue) unfolded spectra of Obs1 and ratio with respect to the best-fit baseline model including the alternative modeling of the ionized emission line using an Compton-thick ionized disc reflection (a), a hot plasma (b) or a Compton-thin photo-ionized emission component (c), respectively. Note that the \emph{Suzaku} XIS-FI and PIN data points have been rebinned to improve the clarity of the plots to a signal-to-noise of 20 and 5, respectively.}
    \end{figure}

\subsection{Ionized emission in Obs1}

The interpretation of the Fe~XXVI Ly$\alpha$ emission line observed in Obs1 is less clear. In \S3.3 we tested three different models and here we discuss in more detail their possible physical origin.

\subsubsection{Ionized accretion disc reflection?}

As reported in \S3.3.1, the ionized emission in Obs1 can be modelled with a disc line profile with $r_\mathrm{in}$$\ge$50~$r_g$. This fit is supported by the spectral variability, which constrain the emission region to be within $\sim 10^3$~$r_g$ from the black hole. However, the successive modelling with more self-consistent ionized accretion disc reflection codes does not provide a satisfactory fit, even allowing for an iron abundance significantly higher than Solar (see Table~3 and panel a of Fig.~4).

Nonetheless, we can not completely rule out this possibility because there could be additional complexities which are difficult to take into account. For instance, Yaqoob et al.~(2003) reported a similar case of detection of Fe~XXV and Fe~XXVI K emission lines from the \emph{Chandra} grating observation of the Seyfert 1 galaxy NGC~7314. The lines appeared variable, slightly redshifted ($cz$$\simeq$1,500~km/s) and unresolved even using the gratings. The authors argued that these lines could indeed originate from a near face-on disc.

In particular, they suggested a series of possible reasons to explain the narrowness of these lines: i) the lines could originate from individual and localized flares or ``hot spots'' on the disc (e.g., Goosmann et al.~2007); ii) a given observed line could be the red or blue Doppler horn of a much wider line profile, originating from a ring/spiral-like region of disc, thereby producing emission lines with sharp ``spikes'' (e.g., Armitage \& Reynolds 2003; Fukumura \& Tsuruta 2004); iii) the entire line profile may come from a large area on the disc but it may be still so narrow due to a flat radial line emissivity and a near face-on configuration. The jet inclination angle of $\theta \sim 18^{\circ}$ does imply a close to face-on geometry. Indeed, this last case is very similar to that suggested by the lamp-post scenario, which provides a best-fit equivalent to the one using a standard relativistic line profile, without requiring a truncated disc. The X-ray source, possibly being represented by the base of the jet, is estimated at a height of $h$$\ga$80~$r_g$ assuming the jet angle or at $h$$\ga$30~$r_g$ for a freely varying inclination angle of $\theta \la 17^{\circ}$.

\begin{table}
\centering
\begin{minipage}{50mm}
\caption{Best-fit parameters of the various models used to characterize the ionized emission line in Obs1.}
\begin{tabular}{lc}
\hline\hline
\multicolumn{2}{l}{(a) Ionized reflection (\emph{Xillver})} \\
\hline
log$\xi$ (erg~s$^{-1}$~cm) & $3.49^{+0.43}_{-0.32}$\\
$A_\mathrm{Fe}$ (Solar) & $>2^a$\\
$N$ ($\times 10^{-9}$) & $2.9^{+2.0}_{-1.3}$\\
$z$ & $0.060^{+0.018}_{-0.014}$\\
\hline
$\chi^2/d.o.f.$ & 2189.5/2145\\
\hline\hline
\multicolumn{2}{l}{(b) Hot plasma (\emph{Apec})} \\
\hline
$kT$ (keV) & $22.0^{+6.1}_{-3.2}$\\
$N$ ($\times 10^{-3}$) & $3.5\pm0.7$\\
$z$ & $0.058\pm0.006$\\
\hline
$\chi^2/d.o.f.$ & 2182.1/2146\\
\hline\hline
\multicolumn{2}{l}{(c) Photo-ionized emission (\emph{Xstar})} \\
\hline
log$\xi$ (erg~s$^{-1}$~cm) & $4.52^{+0.10}_{-0.16}$\\
$N_\mathrm{H}$ ($\times 10^{22}$~cm$^{-2}$) & $>30^a$\\
$N$ ($\times 10^{-4}$) & $\equiv$1.5\\
$z$ & $0.059\pm0.06$\\
\hline
$\chi^2/d.o.f.$ & 2181.1/2146\\
\hline
\end{tabular}
{\em Note.} $^a$ 90\% lower limit. 
\end{minipage}
\end{table}

\subsubsection{ADAF or shock emission?}

As reported in \S3.3.2, a hot plasma component with a very high temperature of $kT$$\simeq$20~keV ($T \simeq 2\times 10^8$~K) can provide a good representation of the emission line as well. A possible interpretation could be a truncated accretion disc with the inner regions replaced by a Compton-thin, hot Advection Dominated Accretion Flow (ADAF; Narayan \& Yi 1995). Thus, the Fe~XXVI emission in Obs1 could originate from ionized material perhaps at the transition region between the hot, inner flow and the cold, truncated accretion disc at $r$$\sim$$10^3$~$r_g$ (e.g., Narayan \& Raymond 1999). Moreover, accretion discs in the ADAF states have been found to produce significant outflows (Blandford \& Begelman 1999; Di Matteo et al.~2003; Tchekhovskoy et al.~2011; Yuan, Bu \& Wu 2012). 

 
The Eddington luminosity of 3C~111 can be estimated as $L_{\mathrm{Edd}}$$=$$1.3\times 10^{38} (M_{\mathrm{BH}}/M_{\odot})$$\simeq$$2\times 10^{46}$~erg~s$^{-1}$. The bolometric luminosity during Obs1 is estimated to be $L_\mathrm{bol}$$\simeq$$6.1\times 10^{45}$~erg~s$^{-1}$ from the relation $L_\mathrm{bol} = k_\mathrm{bol} L_\mathrm{ion}$, where $k_\mathrm{bol}$$\simeq$10 (McKernan et al.~2007; Vasudevan \& Fabian 2009; Lusso et al.~2010), and substituting the ionizing luminosity $L_\mathrm{ion} \simeq 6.1\times 10^{44}$~erg~s$^{-1}$. Therefore, we derive that the Eddington ratio during Obs1 was $\lambda = L_\mathrm{bol}/L_\mathrm{Edd} \simeq$0.1--0.3. This value is much higher than that expected for typical ADAF dominated states, which are $\lambda \la$0.01 (Esin, McClintock \& Narayan 1997). Moreover, there is also recent work showing that the accretion disc might actually be not substantially truncated in hard/intermediate states (Reis et al.~2010, 2011, 2012; Walton et al.~2012; Reynolds \& Miller 2013) and, in the case of the X-ray bright 3C~111, certainly not at at such a large radius of $\sim$10$^3$$r_\mathrm{g}$. Therefore, these considerations strongly disfavour the ADAF interpretation.

From the comparison between the long-term X-ray monitoring of 3C~111 and the radio jet images, Tombesi et al.~(2012b) found that these \emph{Suzaku} observations occurred during a period in which the source was undergoing a significant outburst phase, with the ejection of a superluminal jet knot in the radio and a UFO in the X-rays. This may lead to another possible physical explanation for the origin of this hot plasma component, as a strong shock resulting from the interaction of accretion disc outflows or the jet with the surrounding material. The thermalization of the kinetic energy in a shock derived from an outflow with velocity of $v$$\sim$0.1c would be $kT \simeq (1/3)m_\mathrm{H}v^2$, corresponding to extremely high temperatures of $T$$\sim$$10^{10}$--$10^{11}$~K (e.g., King, Zubovas \& Power 2011; Zubovas \& King 2012). This would be even higher for a relativistic jet. Such temperature is much higher than that estimated from the collisional plasma model and would imply that even iron ions would be completely stripped of their electrons, possibly leaving no observable spectral features, unless the material could cool very rapidly. Moreover, the shock region would correspond to a very narrow shell, which would probably not be extended enough to produce any intense emission features. These considerations also disfavour the possible shock origin.

  \begin{figure}
  \centering
    \includegraphics[width=5.5cm,height=8cm,angle=-90]{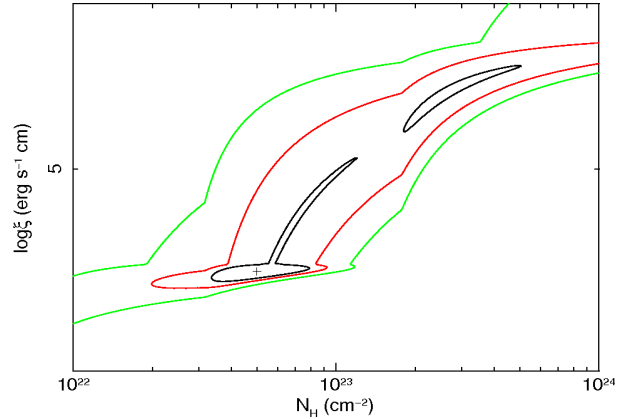}
   \caption{Two-dimensional contour plot of log$\xi$ vs. $N_\mathrm{H}$. The countour levels refer to $\Delta\chi^2$ (confidence) values of 2.3 (black, 68\%), 4.61 (red, 90\%) and 9.21 (green, 99\%) for two additional model parameters. The cross indicates the best-fit solution.}
    \end{figure}

\subsubsection{Photo-ionized emission and absorption from a disc wind?}

Besides Compton-thick reflection from accretion disc material or hot collisional plasma emission, in \S3.3.3 we discussed the possibility to model the line in Obs1 as emission from a Compton-thin, or mildly Compton-thick, photo-ionized plasma (e.g., Lobban et al.~2010; Lohfink et al.~2012). Indeed, the line can be well described with an XSTAR photo-ionized emission component with $N_\mathrm{H} > 3\times 10^{23}$~cm$^{-2}$ and log$\xi$$=$$4.52^{+0.10}_{-0.16}$~erg~s$^{-1}$~cm. This fit provides the lowest $\chi^2/d.o.f$ in Table~3. 

Remarkably, these parameters match those of the $v_\mathrm{out} \simeq 0.1$c UFO in Obs2 (see \S3.4), possibly suggesting a link between these two components. As discussed in \S4.1, substituting the values relative to the photo-ionized emitter and the ionizing luminosity in Obs1 of $L_\mathrm{ion} \simeq 6.1\times 10^{44}$ erg~s$^{-1}$, we derive $r_\mathrm{max} \la 6.1\times 10^{16}$~cm ($\la 3\times 10^3$~$r_g$), which is within that derived for the UFO in \S4.1 using the same formula. This indicates that the emission in Obs1 and absorption in Obs2 could come from material at the same or comparable distance. This possibility is strengthened by the fact that both spectral features are variable on a time-scale of $\sim$7~days.

In order to further investigate the possible link between the emission/absorption phenomena, we performed a test including a photo-ionized emission component also in the spectrum of Obs2, but this time with the column density and ionization parameter linked to the photo-ionized absorption model. Interestingly, we find that the inclusion of the emission component does not significantly change the fit, indicating that for a column density of $N_\mathrm{H} \simeq 5\times 10^{22}$~cm$^{-2}$, derived from the UFO, the emission from the same material would not be detectable. Assuming the same energy as in Obs1, we find a 90\% upper limit for the EW of the ionized emission line in Obs2 of only EW$<$11~eV. The same EW upper limit can be derived for Obs3. This is indeed consistent with the small value of EW$\la$10~eV expected for the emission from a gas with $N_\mathrm{H}$$\sim$1--5$\times 10^{22}$~cm$^{-2}$ (Bianchi \& Matt 2002). Therefore, the data are indeed consistent with a scenario in which the same material could be possibly responsible for both emission in Obs1 and absorption in Obs2. The low EW upper limits of the emission line in Obs2 and Obs3 can be simply explained with a decrease in column density. 

The consistency of both locations and vicinity to the black hole suggest a link with the accretion disc. The decrease of the column density between the photo-ionized emitter and absorber could be due to an expansion of the material outflowing away from the disc. However, the two column density values are essentially consistent considering the larger 90\% errors.  The fact that the emission and absorption lines are detected in two separate observations is not consistent with a scenario in which they are both produced in a fully covering, spherical symmetric wind/outflow because in that case one would expect a P-Cygni line profile, with the simultaneous presence of redshifted emission and blueshifted absorption, which is not observed. This possibility is also hampered by the fact that the opposite/receding part of the outflow would be obscured by the disc itself (e.g., Yaqoob et al.~2003).

Indeed, the photo-ionized Fe K emission/absorption lines could be consistent with an outburst scenario in a conical/bi-polar geometry, which is expected for a wind ejected from a disc (e.g., Proga \& Kallman 2004; Fukumura et al.~2010; Ponti et al.~2012). This possibility is naively illustrated in Fig.~6. In Obs1, the photo-ionized emitting material lies out of the line of sight, possibly close to the accretion disc. In Obs2, the ejected material is seen as an outflowing absorber crossing the line of sight, indicative of a UFO with projected velocity of $v_\mathrm{out} \sim 0.1$c. Instead, in Obs3, the wind is not significantly detected anymore in the spectrum because the column density decreased due to expansion and/or the material moved away from the line of sight.

  \begin{figure}
  \centering
   \includegraphics[width=4.8cm,height=6.5cm,angle=90]{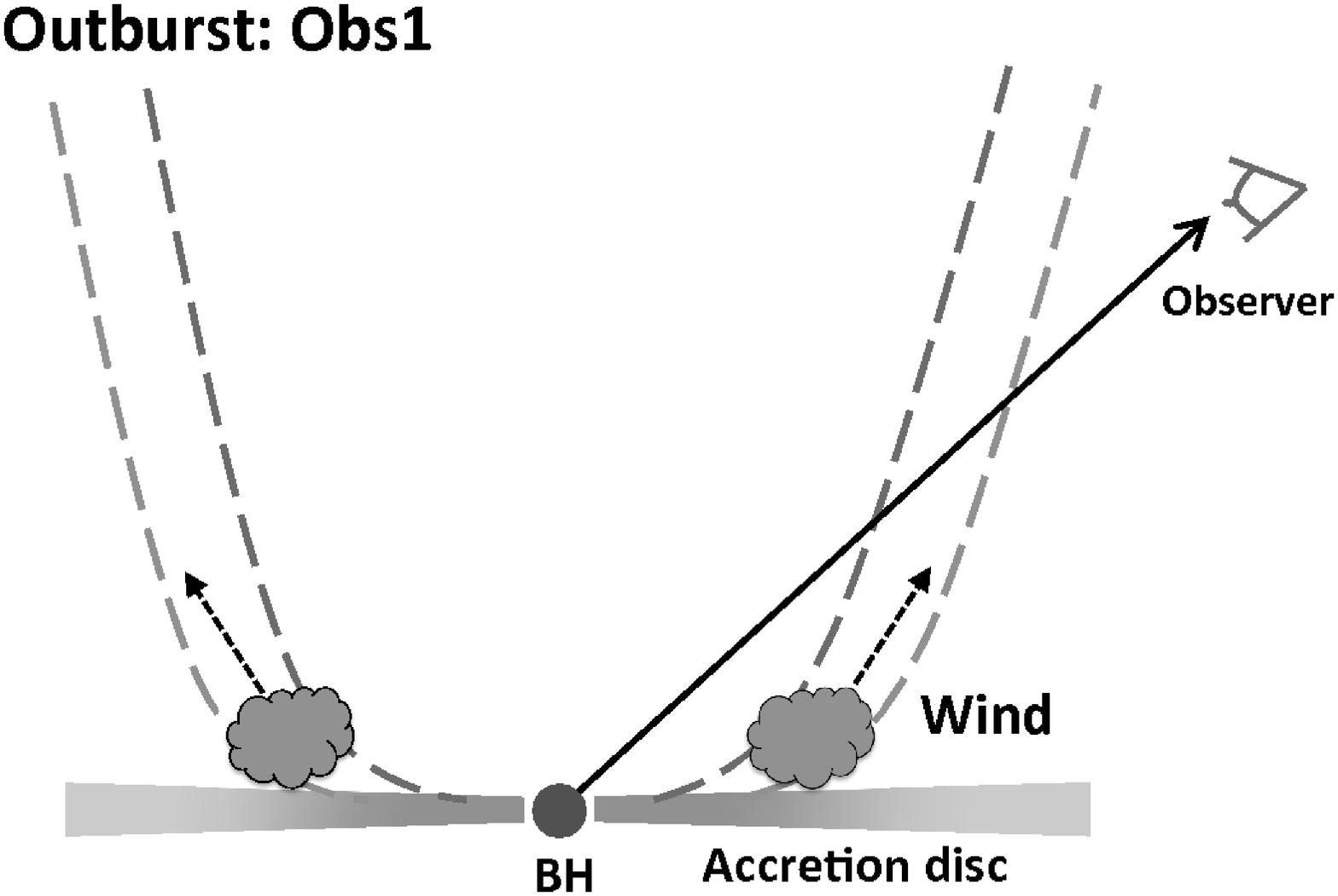}
\hspace{0.5cm}
   \includegraphics[width=4.8cm,height=6.5cm,angle=90]{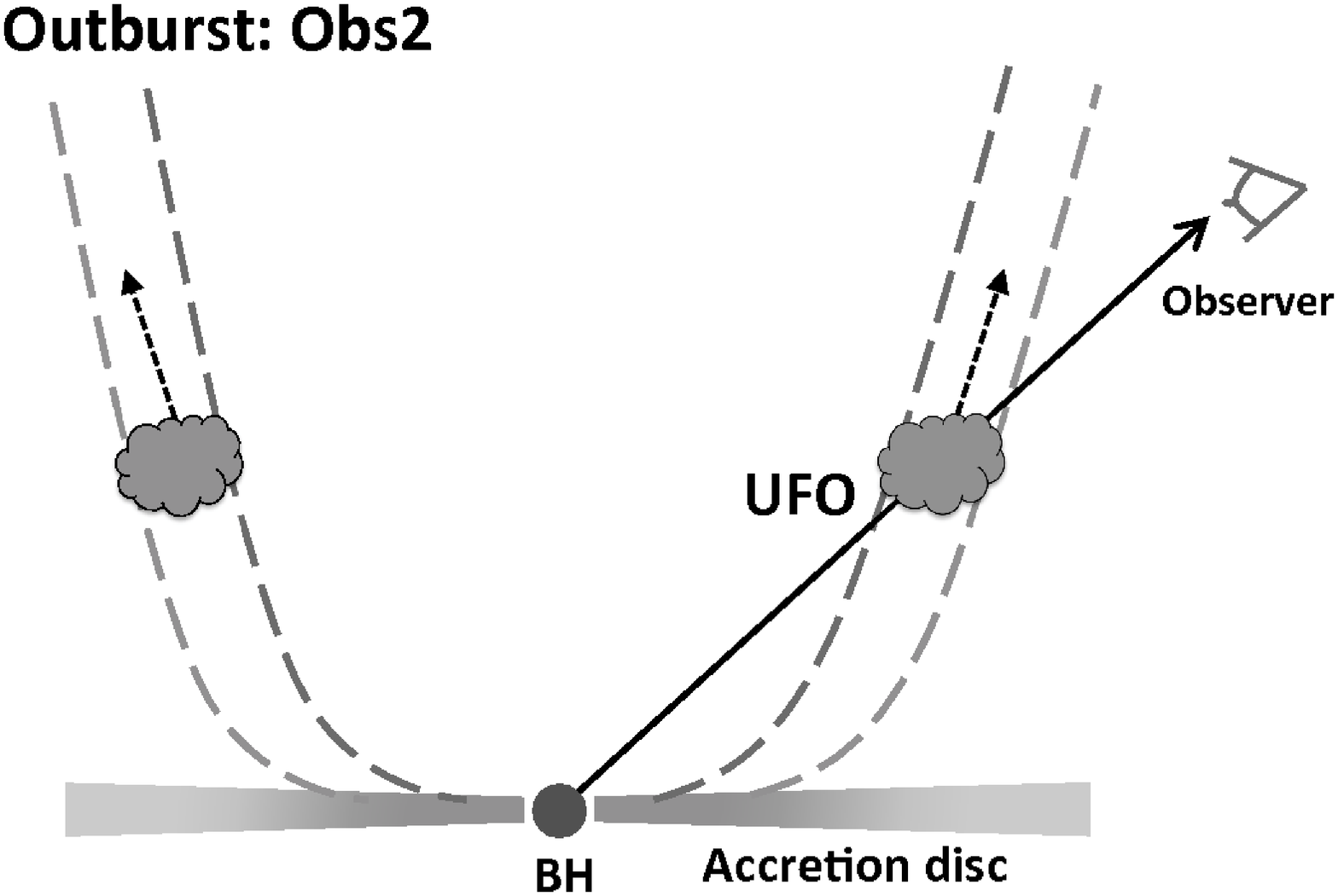}
\hspace{0.5cm}
   \includegraphics[width=4.8cm,height=6.5cm,angle=90]{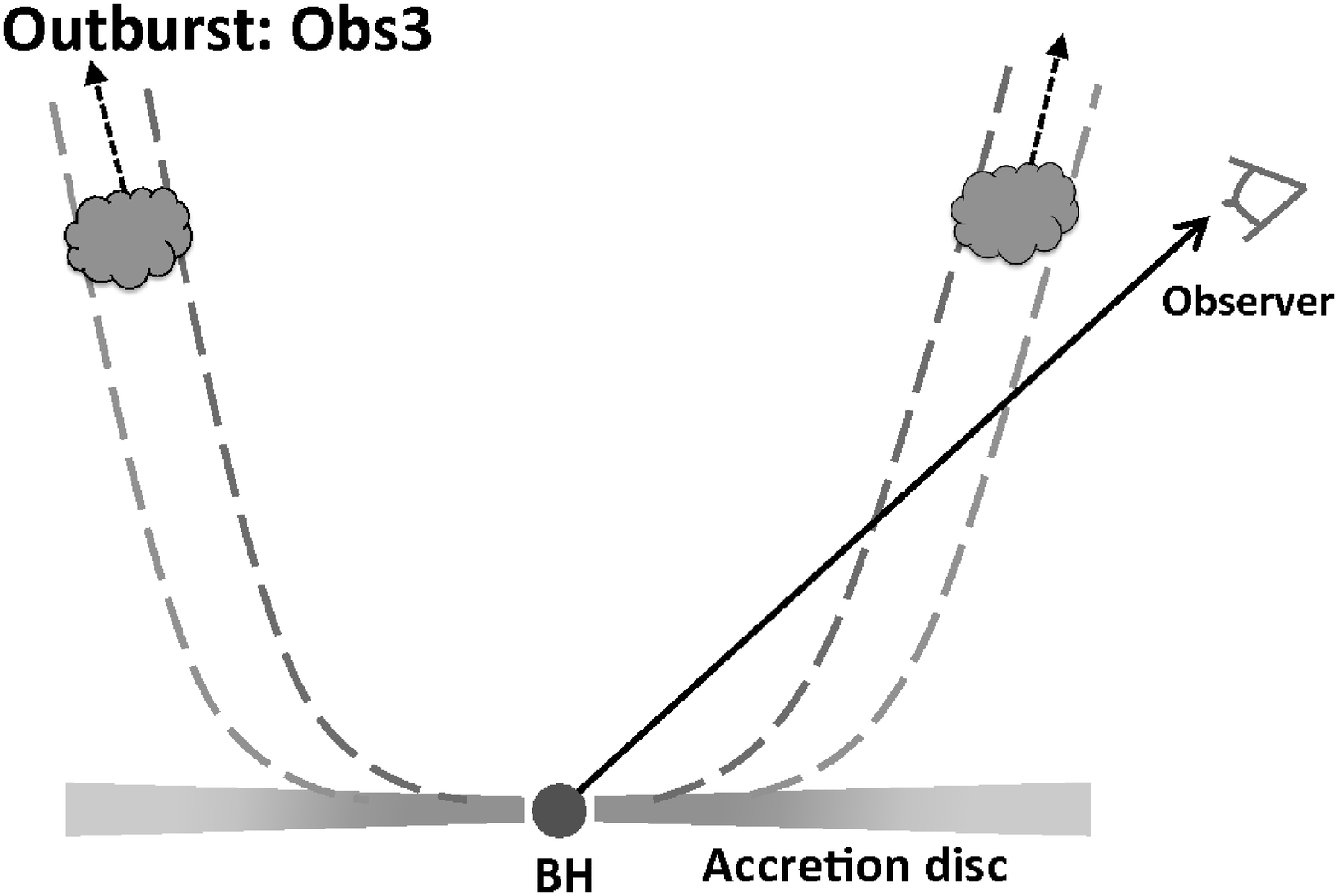}
   \caption{Outburst scenario for the three \emph{Suzaku} observations. In Obs1 the photo-ionized emitting material is out of the line of sight, close to the accretion disc. In Obs2 the material is seen in absorption crossing the line of sight as a UFO with $v_\mathrm{out} \sim 0.1$c. In Obs3 the expanding and outflowing material is not significantly detected anymore because the density decreased too much or it moved out of the line of sight. The figure is not to scale.}
    \end{figure}

As already discussed in Tombesi et al.~(2011b), this suggests that, for the first time, we could be directly witnessing the formation and ejection of an accretion disc outflow or UFO in an AGN. In this respect, the outburst scenario is supported also by the fact that the \emph{Suzaku} observations of 3C~111 performed in September 2010 occurred during a period in which the source was indeed experiencing major jet ejection episodes, which resulted in the formation of new superluminal blobs apparent in the radio jet as well (Tombesi et al.~2012b). The possibility to detect both highly ionized Fe K emission and absorption lines from accretion disc winds in the X-ray spectra of AGNs has been recently reported by several authors (e.g., Sim et al.~2008; 2010a, b; 2012; Pounds \& Reeves 2009; Schurch, Done \& Proga 2009). Similarly to the jet ejection events, the spectral variability suggests a discrete outflow event, rather than a persistent wind, which presumably would result in quasi constant line of sight absorption and emission.

In the outburst or accretion disc wind scenario shown in Fig.~6, the highly ionized material responsible for the photo-ionized emission and absorption would eventually continue its roughly vertical outflowing path, suggesting the possibility that it might give rise to the large-scale ($\gg$1~pc), tenuous electron scattering medium that is observed through polarized light in optical and through soft X-ray emission lines/scattered continuum in type 2 AGNs (e.g., Krolik \& Begelman 1986; Krolik \& Kallman 1987). Here, we do not refer to the compact, X-ray emitting corona, which instead should be localized in the very inner regions of the accretion flow, probably within a few $r_\mathrm{g}$ from the central black hole (Reis \& Miller 2013). In fact, the wind discussed here is also probably ejected from the accretion disc at radii $\ga$100~$r_\mathrm{g}$, therefore much further away than the inner X-ray source.

\section{Conclusions}

In this paper we exploit the unique capabilities of the combined instruments on board \emph{Suzaku} and \emph{Swift} BAT to perform a broad-band spectral analysis of the BLRG 3C~111 in the wide energy range E$=$0.6--200~keV. The data are well described with an absorbed power-law continuum and a weak ($R\simeq 0.2$) cold reflection component from distant material. We constrain the continuum cutoff at $E_\mathrm{C}$$\simeq$150--200~keV, which is in accordance with X-ray Comptonization corona models and supports claims that the jet emission is dominant at radio and $\gamma$-rays. In addition, we detect an emission line at E$\simeq$6.88~keV in Obs1 and an absorption line at E$\simeq$7.75~keV in Obs2, possibly ascribable to highly ionized Fe K features. The latter is well described with a photo-ionized UFO absorber with log$\xi$$=$$4.47^{+0.76}_{-0.04}$~erg~s$^{-1}$~cm, $N_\mathrm{H}$$=$$(5.3^{+1.8}_{-1.3})\times 10^{22}$~cm$^{-2}$ and outflow velocity $v_\mathrm{out} = 0.104\pm0.006$c. It is observable at a location of $r$$\simeq$200--1000~$r_g$ from the SMBH and the mechanical power corresponds to $\simeq$1\% of the bolometric luminosity, consistent with values required for AGN feedback.
Instead, the modelling and interpretation of the Fe XXVI Ly$\alpha$ emission line in Obs1 is more complex and we explore three possibilities. 

\begin{itemize}

\item{} A modelling with a relativistic disc line profile suggests that the emission could originate from the accretion disc at  $r_\mathrm{in} \ge 50$~$r_g$ or, in the lamp-post configuration, that the illuminating source is at a height over the black hole of $h$$\ge$80$r_g$ assuming the jet angle of $\sim18^{\circ}$ or $h$$\ge$30~$r_g$ for a freely varying inclination angle of $\theta\le 17^{\circ}$. over the black hole. However, a successive modelling with more self-consistent ionized accretion disc reflection codes does not provide a satisfactory fit, even allowing for an iron abundance much higher than the standard Solar value. Nonetheless, we can not completely rule out this possibility because there could be additional complexities which are difficult to take into account. 
\item{} An alternative origin from a collisionally ionized plasma requires a very high temperature of $kT = 22.0^{+6.1}_{-3.2}$~keV. This fit is statistically better than the previous one, but it is not clear what would be the physical origin of this hot material. Some possibilities are explored, such as emission from the transition region between an inner, hot ADAF and a cold, truncated accretion disc or a strong shock resulting from outflows/jet. However, both cases are disfavored on physical grounds. 

\item{} Finally, the line could originate from a Compton-thin (or mildly Compton-thick) photo-ionized plasma close to the black hole with log$\xi$$=$$4.52^{+0.10}_{-0.16}$~erg~s$^{-1}$~cm and $N_\mathrm{H} > 3\times 10^{23}$~cm$^{-2}$. Remarkably, these parameters match those of the highly ionized UFO observed in absorption in Obs2. This and other considerations indeed support the possibility that the same material could be possibly responsible for both emission in Obs1 and absorption in Obs2. In Fig.~6 we suggest a possible outburst scenario in which an accretion disc wind, initially lying out of the line of sight, then crosses our view to the source and it is observed as a mildly-relativistic UFO. 

\end{itemize}

The possible link between the ionized emission and absorption features gives rise to the possibility to detect or study AGN accretion disc winds, and in particular UFOs, in emission as well. Therefore, when such a narrow ionized Fe K emission lines are detected, it could suggest that a UFO is present but it is out of our line of sight. Finally, the unprecedented sensitivity and energy resolution in the Fe K band provided by the microcalorimeter onboard the upcoming \emph{Astro-H} mission (Takahashi et al.~2012) will allow to better study these emission/absorption lines.

\section*{Acknowledgments}

The authors thank the referee for the thorough reading and the constructive comments that led to improvements in the paper. FT thank K. Fukumura, D. Kazanas, R. Nemmen, R.~F. Mushotzky, S.~B. Kraemer, M. Cappi and H.~L. Marshall for the useful discussions. FT acknowledges partial support for this work by the National Aeronautics and Space Administration under Grant No. NNX12AH40G issued through the Astrophysics Data Analysis Program, part of the ROSES 2010.

\end{document}